\documentclass[journal]{IEEEtran}

\usepackage{amsmath, amssymb, graphicx, subcaption, cite}
\usepackage[hyphens]{url}

\newtheorem{theorem}{Theorem}

\newtheorem{lemma}{Lemma}
\newtheorem{defn}{Definition}
\newtheorem{corollary}{Corollary}

\DeclareMathOperator*{\argmin}{arg\,min}

\newcommand{\wh}[1]{\widehat{#1}}
\newcommand{\RN}[1]{\textup{\uppercase\expandafter{\romannumeral#1}}}
\newcommand{\calN}{{\mathcal{N}}}
\newcommand{\calL}{{\mathcal{L}}}
\newcommand{\calS}{{\mathcal{S}}}
\newcommand{\pth}[1]{\left( #1 \right)}
\newcommand{\qth}[1]{\left[ #1 \right]}
\newcommand{\sth}[1]{\left\{ #1 \right\}}
\newcommand{\prob}{\mathbb{P}}
\newcommand{\Prob}{\mathbb{P}}

\title{Beliefs in Decision-Making Cascades}

\author{Daewon Seo, Ravi Kiran Raman, Joong Bum Rhim, Vivek K Goyal, Lav R.~Varshney
	\thanks{ This paper was presented in part at the IEEE International Symposium on Information Theory (ISIT 2013) \cite{RhimG2013} and at the IEEE International Conference on Acoustics, Speech and Signal Processing (ICASSP 2018) \cite{SeoRV2018}. This work was supported in part by the National Science Foundation under Grants CCF-1101147 and CCF-1717530. }
	\thanks{ D.~Seo, R.~K.~Raman, and L.~R.~Varshney are with the Coordinated Science Laboratory, and the Department of Electrical and Computer Engineering, University of Illinois at Urbana-Champaign, Urbana, IL 61801 USA (e-mail: \{dseo9, rraman10, varshney\}@illinois.edu). }
	\thanks{ J.~B.~Rhim is with GroupM, New York, NY 10007 USA (e-mail: jbrhim@alum.mit.edu). }
	\thanks{ V.~K.~Goyal is with the Department of Electrical and Computer Engineering, Boston University, Boston, MA 02215 USA (e-mail: v.goyal@ieee.org). }
}
\date{}

\begin{document}
\maketitle
	
\begin{abstract}
This work explores a social learning problem with agents having nonidentical noise variances and mismatched beliefs. We consider an $N$-agent binary hypothesis test in which each agent sequentially makes a decision based not only on a private observation, but also on preceding agents' decisions. In addition, the agents have their own beliefs instead of the true prior, and have nonidentical noise variances in the private signal. We focus on the Bayes risk of the last agent, where preceding agents are selfish.

We first derive the optimal decision rule by recursive belief update and conclude, counterintuitively, that beliefs deviating from the true prior could be optimal in this setting. The effect of nonidentical noise levels in the two-agent case is also considered and analytical properties of the optimal belief curves are given. Next, we consider a predecessor selection problem wherein the subsequent agent of a certain belief chooses a predecessor from a set of candidates with varying beliefs. We characterize the decision region for choosing such a predecessor and argue that a subsequent agent with beliefs varying from the true prior often ends up selecting a suboptimal predecessor, indicating the need for a social planner. Lastly, we discuss an augmented intelligence design problem that uses a model of human behavior from cumulative prospect theory and investigate its near-optimality and suboptimality.

\end{abstract}


\section{Introduction} \label{sec:intro}
Team decision-making typically involves individual decisions that are influenced by private observations and the opinions of the rest of the team. The \emph{social learning} setting is one such context where decisions of individual agents are influenced by preceding agents in the team \cite{EllisonF1993, KrishnamurthyP2013}. We consider the setting in which individual agents are selfish and aim to minimize their perceived Bayes risk, according to their beliefs as reinforced by the decisions of preceding agents. 

Social learning, also referred to as observational learning, has been widely studied and we provide a non-exhaustive listing of some of the relevant works. Aspects of conformism and ``herding'' were studied in \cite{Banerjee1992, BikhchandaniHW1998, BalaG2001}, where an incorrect decision may cascade for the rest of the agents once agents at the beginning make incorrect decisions. The concept of herding is a consequence of boundedly informative private signals \cite{SmithS2000}. For example, assume the private signals are binary and give true or false information, each with positive probability. It can happen that a couple of the first agents receive false private signals and thus choose wrong actions. Then, the effect of these actions on the beliefs of subsequent agents can be so great as to cause them to ignore their private signals and follow their predecessors. The private signals are bounded so that they are not strong enough to overcome the effect of the wrong actions. Further convergence properties of actions taken under social learning have been explored under imperfect information \cite{CelenK2004}. The notion of sequential social learning has been generalized to learning from neighbors in networks \cite{GaleK2003}, and explored in generality \cite{AcemogluDLO2011}. Social learning has also been explored under quantization of priors \cite{RhimVG2012}, and distributed detection with symmetric fusion \cite{RhimG2014}.

Such social learning problems have also been studied as \emph{distributed inference} or \emph{learning}. The traditional setting assumes a central fusion node that aggregates all information from distributed nodes and makes the final decision \cite{VeeravalliBP1993, ViswanathanV1997}, where the links between distributed nodes and fusion center could be rate-limited \cite{BergerZV1996} or imperfect \cite{SaligramaAS2006, KarM2009, KarM2010}. It is also common to consider such learning over networks. The network setting could be the simplest tandem network (in particular, this is called serial detection) \cite{HellmanC1970, TangPK1991, PapastavrouA1992, TayTW2008} and extended to a general network, in which all nodes can identify the hypothesis by repeatedly updating local beliefs without complete knowledge of network connectivity \cite{AlanyaliVSA2004, RadT2010, JadbabaieMST2012}. Independent works \cite{ShahrampourRJ2016} and \cite{NedicOU2015} propose similar update rules and convergence results for fixed networks and time-varying networks, respectively. In \cite{SahuK2016}, binary hypothesis testing in the presence of Gaussian process noise is studied and minimal expected stopping times are derived. In \cite{LalithaJS2018}, the setup where the entire hypotheses are locally indistinguishable, but globally identifiable by belief update is considered and convergence rate is provided.

This paper differs from the literature in the sense that we consider unbounded private signals so that there is no herding behavior. Unlike typical decision-making cascades (e.g., \cite{Varshney1997}) where all agents know the true prior, we assume agents may have \emph{beliefs} that do not necessarily match the true prior. Further, private signal strengths of agents could be different, i.e., noise variances are not necessarily identical. Information is only propagated along the chain once so there is no iterative belief update. We focus largely on the effects of initial belief and private signal strength. The decision-making of individual agents is also different in that agents make locally Bayes-optimal decisions, i.e., decisions that minimize their individual Bayes risk. This is different from the context of collectively optimizing the team's risk \cite{RhimG2014} or decision-making that maximizes a personal reward based on the last agent's decision \cite{RamanP2018}.

We study cascading binary hypothesis testing (or sequential social learning following the notion of \cite{GaleK2003}) and characterize optimal beliefs of agents that minimize the Bayes risk of the last-acting agent. In general, it counterintuitively turns out that agents using beliefs that do not match the true prior are optimal, i.e., each agent has a perceived \emph{belief} of the prior. For instance, in the two-agent system with equal noise levels, the optimal predecessor is one who overweights the belief for small prior, and underweights when it is large. On the other hand, the corresponding optimal last agent is one who behaves in the opposite way to the predecessor. For concise description, we refer to these two modes of operation as \emph{open-minded} and \emph{closed-minded}, respectively.\footnote{As far as we know, these terms were first introduced in \cite{Radner1962}.} We describe analytical characteristics of the optimal beliefs and also show how the nature of such behaviors of agents change when noise levels differ in the private signals.

We are ultimately interested in the Bayes risk of the last-acting agent, and thus it is important that the last agent uses the correct set of preceding agents for the task. To this end, we consider a team construction problem for such cascading hypothesis testing, and characterize the criterion used for predecessor selection. We observe that self-organized teams may have suboptimal compositions, emphasizing the importance of a social planner that is aware of the true prior. 

We also consider a collaborative decision-making system with human and AI (Artificial Intelligence or Augmented Intelligence). A cascading decision-making model captures the nature of collaboration in human-AI teams with either the AI system advising the human who makes the final decision or less typically a human advising an AI system that makes the final decision \cite[p.~56]{McAfeeB2017}. Examples of the first kind include AI-assisted physicians, and of the second kind, human-in-the-loop AI systems such as crowdsourcing.

Note that in human-AI systems, human actions are affected by individual perceptions of the underlying context that cannot be tuned/controlled like machine agents (e.g.\ sensor nodes). Cumulative prospect theory \cite{TverskyK1992, NadendlaBV2016, NadendlaALB2017} seeks to describe boundedly rational human behavior under risk by introducing probability reweighting functions. Among reweighting functions, the Prelec reweighting function \cite{Prelec1998} has significant empirical support and satisfies a majority of the axioms of prospect theory. We first show the Prelec model does not capture all patterns of optimal beliefs of agents in the case of diverse noise levels.  Next we show that a team of suboptimal human-AI agents could outperform a team of standalone optimal human-AI agents, if it is well-composed.

The rest of this paper is organized as follows. Sec.~\ref{sec:prob_desc} describes the cascading binary hypothesis testing problem. Sec.~\ref{sec:belief_update} proposes a recursive belief update equation that transforms the cascading hypothesis testing problem into a single-agent binary hypothesis testing problem. Sec.~\ref{sec:optimal_belief} shows the optimal beliefs that minimize the last agent's Bayes risk and Sec.~\ref{sec:Gaussian_diverse_expertise} evaluates them for Gaussian likelihoods. Sec.~\ref{sec:team_construction} considers a two-agent team construction problem and Sec.~\ref{sec:human_ai} discusses design principles of AI-human collaboration systems. Sec.~\ref{sec:conclusion} concludes.

This cascading decision-making problem with identically noisy agents was first presented in \cite{RhimG2013} and in particular two-agent systems with varying noise levels were investigated in \cite{SeoRV2018}. This paper integrates and generalizes our previous results, and also, significantly improves analytic understanding on optimal beliefs in cascading decision-making. In addition, we provide a novel interpretation of Prelec-like beliefs in terms of AI-human collaboration systems.

\section{Problem Description} \label{sec:prob_desc}
\begin{figure}[t]
	\centering
	\includegraphics[width=3.5in]{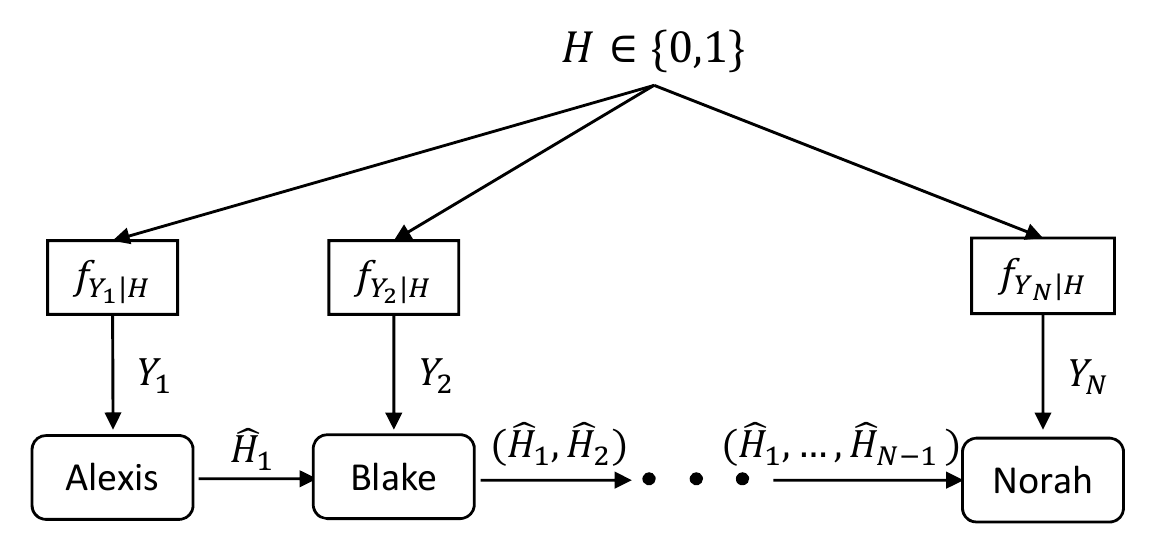}
	\caption{A cascading decision making model with $N$ agents.}
	\label{fig:general_n_user}
\end{figure}

Consider an $N$-agent cascading decision making problem, as illustrated in Fig.~\ref{fig:general_n_user}. The underlying hypothesis, $H \in \{0,1\}$, is a binary signal with prior $\prob{H=0} = p_0$ and $\prob{H = 1} = 1 - p_0$. There are $N$ agents that sequentially detect the state in a predetermined order. The $n$th agent has a \emph{private} signal $Y_n$ generated according to the likelihood $f_{Y_n|H}$, which is not necessarily identical for all $n$. Let the decision made by the $n$th agent be $\wh{H}_n$. In addition to the private signal, the $n$th agent also observes the decisions made by preceding agents, $\{\wh{H}_1, \ldots, \wh{H}_{n-1}\}$, to make a decision $\wh{H}_n$.

However, the $n$th agent believes the prior probability of the null hypothesis is $q_n \in (0,1)$ as against the true prior probability $p_0$. We call this the \emph{belief} of the agent in order to distinguish it from the prior. Agent $n$ is also aware of her own likelihood $f_{Y_n|H}$ that defines her private signal. However, she also perceives the likelihoods and beliefs of the other agents to be the same as hers, i.e., she thinks $f_{Y_j|H} = f_{Y_n|H}, q_j = q_n$ for all $j \neq n$, even though they could be different and unknown to her. We assume that the likelihood ratio of each agent is an increasing function in $y$,\footnote{This property is particularly useful in uniformly most powerful (UMP) tests.} i.e., for all agents
\begin{align*}
	\mathcal{L}_n(y) := \frac{f_{Y_n|H}(y|1)}{f_{Y_n|H}(y|0)}
\end{align*}
is an increasing function of $y$.

Several numerical examples are given for private signals defined with independent additive Gaussian noise. The desired monotonicity also holds for many non-additive models, such as exponential distribution with mean $H^{-1}$, $H \in \mathbb{R}_+$, binomial distribution with success probability $H \in [0,1]$, and Poisson distribution with rate $H \in \mathbb{R}_+$ are members of such family, where $H$ could take two values.

Our performance analysis focuses on the last agent ($N$th agent, Norah) and her decision ̂$\wh{H}_N$. Upon observing her private signal $Y_N$ and the $(N - 1)$ preceding decisions, she determines her decision rule. The relative importance of correct decisions and errors can be abstracted as a cost function. For simplicity, we assume correct decisions have zero cost and use the shorthand notations $c_{10} = c(1, 0)$ as the cost for false alarm or Type I error (choosing ̂$\wh{H}=1$ when $H = 0$), and $c_{01} = c(0, 1)$ as the cost for missed detection or Type II error (choosing $\wh{H}=0$ when $H = 1$). In addition, we assume that agents have the same costs; they are a team in the sense of Radner \cite{Radner1962}. Then the Bayes risk is
\begin{align}
	R_N = c_{10} p_0 p_{\wh{H}_N|H}(1|0) + c_{01} (1-p_0) p_{\wh{H}_N|H}(0|1). \label{eqn:Bayes_risk_n}
\end{align}
As $\wh{H}_n$ depends on the previous decisions, the computation of \eqref{eqn:Bayes_risk_n} also depends on $(\wh{H}_1, \ldots, \wh{H}_{N-1})$, and the Bayes risk can be expanded as
\begin{align}
	R_N &= \sum_{ \wh{h}_1, \ldots, \wh{h}_{N-1}} c_{10} p_0 p_{\wh{H}_N, \wh{H}_{N-1}, \ldots, \wh{H}_1|H}(1, \wh{h}_{N-1}, \ldots, \wh{h}_1|0) \nonumber \\
	&+ c_{01} (1-p_0) p_{\wh{H}_N, \wh{H}_{N-1}, \ldots, \wh{H}_1|H}(0, \wh{h}_{N-1}, \ldots, \wh{h}_1|1).  \label{eqn:Bayes_risk_expansion}
\end{align}
We determine the optimal set of beliefs of the agents $\{q_n^*\}_{n=1}^N$ that minimize \eqref{eqn:Bayes_risk_expansion}.

In our model, the $n$th agent minimizes her \emph{perceived} Bayes risk, which is the Bayes risk with prior probability $p_0$ replaced by her belief $q_n$. In other words, for all $n=1, \ldots, N$, the $n$th agent adopts the decision rule that minimizes her perceived Bayes risk $R_n$, and her decision is revealed to other agents as a public signal. The decisions $\{\wh{H}_1, \ldots, \wh{H}_{n-1}\}$ of the earlier-acting agents reveal information about $H$ and thus should be incorporated into the decision-making process by agent $n$. As mentioned earlier, since she believes $q_n$ is the true prior, she aggregates information under the assumption that $q_1 = q_2 = \cdots = q_n$.

It is important to note that every agent is selfish and rational; the agents do not adjust their decision rules for Norah's sake. The novelty in the model (and hence in the conclusions) comes from agent $n$ having the limitation of using a private initial belief $q_n$ in place of the true prior probability $p_0$.

\subsection{Prospect Theory}

Let us also formally introduce the Prelec reweighting function from cumulative prospect-theoretic models of human behavior. It spans a family of open- and closed-minded beliefs (will be clarified later) and thus the optimal beliefs that emerge in following sections could be approximated by a function in the Prelec family.
\begin{defn}[\cite{Prelec1998}] \label{def:prelec}
	For $\alpha, \beta > 0$, the Prelec reweighting function $w:[0,1] \mapsto [0,1]$ is 
	\begin{align*}
	w(p;\alpha,\beta) = \exp(-\beta(-\log p)^{\alpha}).
	\end{align*}
\end{defn}
The function satisfies several properties such as: 
\begin{enumerate}
	\item $w(p;\alpha, \beta)$ is strictly increasing;
	\item has a unique fixed point $w(p;\alpha, \beta) = p$ at $p^* = \exp( -\exp( \log\beta / (1-\alpha)))$; and
	\item spans a class of open-minded beliefs when $\alpha < 1$, i.e., overweights (underweights) small (high) probability, and vice versa when $\alpha > 1$.
\end{enumerate}
A more generic form, termed composite Prelec weighting function, has been defined in \cite{al-NowaihiD2010}.

\subsection{Notations}

Throughout the paper, we use $f$ for continuous probability density functions and $p$ for discrete probability mass functions. All logarithms are natural logarithms. We use $\mathcal{N}(\mu, \sigma^2)$ to denote a Gaussian distribution with mean $\mu$ and variance $\sigma^2$, and $\phi(x; \mu, \sigma^2)$ to denote its density function, i.e.,
\begin{align*}
\phi(x; \mu, \sigma^2) = \frac{1}{\sqrt{2\pi \sigma^2}} e^{-\frac{(x-\mu)^2}{2\sigma^2}}.
\end{align*}
Also in the case of the standard Gaussian, $\phi(x) := \phi(x; 0, 1)$ for simplicity. $Q(x)$ is defined as the complementary cumulative distribution function of the standard Gaussian,
\begin{align*}
Q(x) = \int_{x}^{\infty} \phi(t) dt.
\end{align*}

\section{Belief Update and Sequential Decision Making} \label{sec:belief_update}

Our model assumes unbounded private signals. Thus, unlike in \cite{Banerjee1992,BikhchandaniHW1998}, it is always possible that a subsequent agent may not follow previous decisions; that is, herding happens with arbitrarily low probability. We now discuss using both a decision history and private signals for Bayesian binary hypothesis testing. The decision rule can be interpreted as each agent updating her posterior belief based on the decision history and then applying a likelihood ratio test to her private signal.

\subsection{Alexis, the First Agent}

Since Alexis has no prior decision history, she follows usual binary hypothesis testing. She uses the following likelihood ratio test with her initial belief $q_1$, with ties broken arbitrarily:
\begin{align}
	\mathcal{L}_1(y_1) = \frac{f_{Y_1|H}(y_1|1)}{f_{Y_1|H}(y_1|0)} \underset{\wh{H}_1 = 0}{\overset{\wh{H}_1 = 1}{\gtrless}} \frac{c_{10}q_1}{c_{01}(1-q_1)}. \label{eqn:likelihood_Alexis}
\end{align}
Since we assume the likelihood ratio is increasing in $y_1$, the rule simplifies to comparing the private signal with an appropriate decision threshold:
\begin{align}
	y_1 \underset{\wh{H}_1 = 0}{\overset{\wh{H}_1 = 1}{\gtrless}} \lambda_1(q_1), \label{eqn:decision_threshold}
\end{align}
where $\lambda_i(q)$ denotes the decision threshold $\lambda$ that satisfies
\begin{align}
	\mathcal{L}_i(\lambda) = \frac{f_{Y_i|H}(\lambda|1)}{f_{Y_i|H}(\lambda|0)} = \frac{c_{10}q}{c_{01}(1-q)}. \label{eqn:likelihood_test}
\end{align}

\subsection{Blake, the Second Agent}

Blake observes Alexis's decision $\wh{H}_1 = \wh{h}_1$ and evaluates the likelihood ratio for $(\wh{H}_1, Y_2)$, using his initial belief $q_2$ as
\begin{align}
	\frac{f_{Y_2, \wh{H}_1|H}(y_2,\wh{h}_1|1)}{f_{Y_2,\wh{H}_1|H}(y_2,\wh{h}_1|0)} \underset{\wh{H}_2 = 0}{\overset{\wh{H}_2 = 1}{\gtrless}} \frac{c_{10}q_2}{c_{01}(1-q_2)}. \label{eqn:likelihood_Blake}
\end{align}
The private signals $Y_1$ and $Y_2$ are independent conditioned on $H$, so $\wh{H}_1$ and $Y_2$ are also independent conditioned on $H$. Hence, the left side of \eqref{eqn:likelihood_Blake} is
\begin{align*}
	f_{Y_2, \wh{H}_1|H}(y_2,\wh{h}_1|h) = f_{Y_2|H}(y_2|h) p_{\wh{H}_1|H}(\wh{h}_1|h).
\end{align*}
So we can rewrite \eqref{eqn:likelihood_Blake} as\footnote{The subscript $[2]$ in the term $p_{\wh{H}_1|H}(\wh{h}_1|h)_{[2]}$ indicates the value of $p_{\wh{H}_1|H}(\wh{h}_1|h)$ that Blake (the second agent) thinks. We specify this because Blake does not know Alexis's belief $q_1$. Thus, he interprets her decision based on his belief $q_2$. The value is different from the true value of $p_{\wh{H}_1|H}(\wh{h}_1|h) = p_{\wh{H}_1|H}(\wh{h}_1|h)_{[1]}$. Of course, it will also be different from what Chuck, the third agent, perceives, which is denoted by $p_{\wh{H}_1|H}(\wh{h}_1|h)_{[3]}$. This will be explained in the next subsection.}
\begin{align}
	\frac{f_{Y_2|H}(y_2|1)}{f_{Y_2|H}(y_2|0)} \underset{\wh{H}_2 = 0}{\overset{\wh{H}_2 = 1}{\gtrless}} \frac{c_{10}q_2}{c_{01}(1-q_2)} \frac{p_{\wh{H}_1|H}(\wh{h}_1|0)_{[2]}}{p_{\wh{H}_1|H}(\wh{h}_1|1)_{[2]}}. \label{eqn:likelihood_Blake_2}
\end{align}

The likelihood ratio test \eqref{eqn:likelihood_Blake_2} can be interpreted as Blake updating his initial belief upon observing Alexis's decision $\wh{H}_1$. Combined with $q_2$, his initial belief is updated according to $p_{\wh{H}_1|H}(\wh{h}_1|h)_{[2]}$, from $q_2$ to $q_2^{\wh{h}_1}$:
\begin{align}
	\frac{q_2^{\wh{h}_1}}{1-q_2^{\wh{h}_1}} = \frac{q_2}{1-q_2} \frac{p_{\wh{H}_1|H}(\wh{h}_1|0)_{[2]}}{p_{\wh{H}_1|H}(\wh{h}_1|1)_{[2]}}. \label{eqn:Blake_belief_update}
\end{align}
The posterior belief is
\begin{equation} \label{eqn:belief_update1}
\begin{aligned}
	q_2^{\wh{h}_1} &= \frac{q_2 p_{\wh{H}_1|H}(\wh{h}_1|0)_{[2]} }{q_2 p_{\wh{H}_1|H}(\wh{h}_1|0)_{[2]} + (1-q_2) p_{\wh{H}_1|H}(\wh{h}_1|1)_{[2]}} \\
	&= \frac{p_{\wh{H}_1,H}(\wh{h}_1,0)_{[2]}}{ p_{\wh{H}_1,H}(\wh{h}_1,0)_{[2]} + p_{\wh{H}_1,H}(\wh{h}_1,1)_{[2]} } \\
	&= p_{H|\wh{H}_1}(0|\wh{h}_1)_{[2]}. 
\end{aligned}
\end{equation}

It should be noted that the true $p_{\wh{H}_1|H}(\wh{h}_1|h)$ is given by
\begin{align*}
	p_{\wh{H}_1|H}(0|h) &= p_{\wh{H}_1|H}(0|h)_{[1]} = \prob{Y_1 \le \lambda_1(q_1)|H=h} \\
	&= \int_{-\infty}^{\lambda_1(q_1)} f_{Y_1|H}(y|h) dy, \\
	p_{\wh{H}_1|H}(1|h) &= \int_{\lambda_1(q_1)}^{\infty} f_{Y_1|H}(y|h) dy.
\end{align*}
But Blake evaluates Alexis's decision $\wh{H}_1$ as if it were made based on $q_2$ and the likelihood $f_{Y_2\vert H}(\cdot)$, as against $q_1, f_{Y_1\vert H}(\cdot)$ respectively. Thus the probability $p_{\wh{H}_1|H}(\wh{h}_1|h)$ is computed based on $\lambda_2(q_2)$, instead of $\lambda_1(q_2)$:
\begin{subequations} \label{eqn:belief_update3}
\begin{align}
	p_{\wh{H}_1|H}(0|h)_{[2]} &= \int_{-\infty}^{\lambda_2(q_2)} f_{Y_2|H}(y|h) dy, \\
	p_{\wh{H}_1|H}(1|h)_{[2]} &= \int_{\lambda_2(q_2)}^{\infty} f_{Y_2|H}(y|h) dy.
\end{align}
\end{subequations}

An interesting observation is that Alexis's belief $q_1$ does not affect Blake's belief update as observed in \eqref{eqn:belief_update1} and \eqref{eqn:belief_update3}. That is, for any belief $q_1$ that Alexis might hold, Blake, who does not know this belief, presumes that the conditional probabilities are computed according to \eqref{eqn:belief_update3} and updates his belief as in \eqref{eqn:belief_update1} which depends only on Blake's initial belief and Alexis's decision.

However, Alexis's initial belief implicitly affects Blake's performance since her biased belief changes the resulting decisions whose probabilities are embedded in the probability of Blake's decision:
\begin{align*}
	p_{\wh{H}_2|H}(\wh{h}_2|h) &= \sum_{\wh{h}_1 \in \{0,1\}} p_{\wh{H}_2, \wh{H}_1|H}(\wh{h}_2, \wh{h}_1|h) \\
	&= p_{\wh{H}_2|\wh{H}_1,H}(\wh{h}_2|0,h)_{[2]} \times p_{\wh{H}_1|H}(0|h)_{[1]} \\
	&~~~ + p_{\wh{H}_2|\wh{H}_1,H}(\wh{h}_2|1,h)_{[2]} \times p_{\wh{H}_1|H}(1|h)_{[1]}.
\end{align*}
Thus, Alexis's biased belief changes the probability of not only her decision but also of Blake's decision.

\subsection{Chuck, the Third Agent }

Chuck's detection process is similar to Blake's. He observes both Alexis's and Blake's decisions and also updates his initial belief $q_3$ like in \eqref{eqn:Blake_belief_update}:
\begin{equation} \label{eqn:Chuck_update_eq}
\begin{aligned}
	\frac{q_3^{\wh{h}_1, \wh{h}_2}}{1-q_3^{\wh{h}_1, \wh{h}_2}} &= \frac{q_3}{1-q_3} \frac{p_{\wh{H}_2, \wh{H}_1|H}(\wh{h}_2, \wh{h}_1|0)_{[3]}}{p_{\wh{H}_2, \wh{H}_1|H}(\wh{h}_2, \wh{h}_1|1)_{[3]}} \\
	&= \left( \frac{q_3}{1-q_3} \frac{p_{\wh{H}_1|H}(\wh{h}_1|0)_{[3]}}{p_{\wh{H}_1|H}(\wh{h}_1|1)_{[3]}} \right) \frac{p_{\wh{H}_2|\wh{H}_1,H}(\wh{h}_2|\wh{h}_1, 0)_{[3]}}{p_{\wh{H}_2 | \wh{H}_1,H}(\wh{h}_2|\wh{h}_1,1)_{[3]}}. 
\end{aligned}
\end{equation}
Note that $\wh{H}_1$ and $\wh{H}_2$ are not conditionally independent given $H$ as Blake's decision $\wh{H}_2$ depends on Alexis's decision $\wh{H}_1$.

Chuck's belief update can be understood as a two-step process. The first step is to update his belief according to Alexis's decision:
\begin{align}
	\frac{q_3^{\wh{h}_1}}{1-q_3^{\wh{h}_1}} = \frac{q_3}{1-q_3} \frac{p_{\wh{H}_1|H} (\wh{h}_1|0)_{[3]}}{p_{\wh{H}_1|H} (\wh{h}_1|1)_{[3]}}. \label{eqn:Chuck_updata1}
\end{align}
The second step is to update it from $q_3^{\wh{h}_1}$ based on Blake's decision:
\begin{align}
	\frac{q_3^{\wh{h}_1, \wh{h}_2}}{1-q_3^{\wh{h}_1, \wh{h}_2}} = \frac{q_3^{\wh{h}_1}}{1-q_3^{\wh{h}_1}} \frac{p_{\wh{H}_2|\wh{H}_1,H} (\wh{h}_2|\wh{h}_1,0)_{[3]}}{p_{\wh{H}_2|\wh{H}_1,H} (\wh{h}_2|\wh{h}_1,1)_{[3]}}. \label{eqn:Chuck_updata2}
\end{align}
Again, Chuck is not aware of neither Alexis's nor Blake's initial beliefs or likelihoods. Thus, Chuck computes all probabilities based on his own belief $q_3$ and likelihood $f_{Y_3|H}$, which is indicated by the subscript $[3]$ in \eqref{eqn:Chuck_updata1} and \eqref{eqn:Chuck_updata2}.

Details of computations of \eqref{eqn:Chuck_updata1} and \eqref{eqn:Chuck_updata2} are as follows:
\begin{align*}
	p_{\wh{H}_1|H}(0|h)_{[3]} &= \int_{-\infty}^{\lambda_3(q_3)} f_{Y_3|H}(y|h)dy, \\
	p_{\wh{H}_1|H}(1|h)_{[3]} &= \int_{\lambda_3(q_3)}^{\infty} f_{Y_3|H}(y|h)dy.
\end{align*}
Similar to Blake \eqref{eqn:Blake_belief_update}, Chuck computes $q_3^{\wh{h}_1}$ for $\wh{H}_1=0$ and $\wh{H}_1=1$ respectively as:
\begin{subequations} \label{eqn:Chuck_belief_update}
\begin{align}
	q_3^0 &= \frac{q_3}{q_3 + (1-q_3) \frac{\int_{-\infty}^{\lambda_3(q_3)} f_{Y_3|H}(y|1)dy}{\int_{-\infty}^{\lambda_3(q_3)} f_{Y_3|H}(y|0)dy}}, \\
	q_3^1 &= \frac{q_3}{q_3 + (1-q_3) \frac{\int_{\lambda_3(q_3)}^{\infty} f_{Y_3|H}(y|1)dy}{\int_{\lambda_3(q_3)}^{\infty} f_{Y_3|H}(y|0)dy}}.
\end{align}
\end{subequations}
Then,
\begin{subequations} \label{eqn:Chuck_final_eq}
\begin{align}
	p_{\wh{H}_2|\wh{H}_1, H}(0|\wh{h}_1, h)_{[3]} &= \int_{-\infty}^{\lambda_3(q_3^{\wh{h}_1})} f_{Y_3|H}(y|h)dy, \\
	p_{\wh{H}_2|\wh{H}_1, H}(1|\wh{h}_1, h)_{[3]} &= \int_{\lambda_3(q_3^{\wh{h}_1})}^{\infty} f_{Y_3|H}(y|h)dy.
\end{align}
\end{subequations}
Even though the value of $\wh{h}_1$ does not appear in \eqref{eqn:Chuck_final_eq}, it is implicit in $q_3^{\wh{h}_1}$ and affects the computation results. Chuck's posterior belief $q_3^{\wh{h}_1, \wh{h}_2}$ is obtained by substituting \eqref{eqn:Chuck_belief_update} and \eqref{eqn:Chuck_final_eq} in \eqref{eqn:Chuck_updata2}.

\subsection{Norah, the $N$th Agent}

Norah, the $N$th agent, observes $Y_N$ and $\{\wh{H}_1, \ldots, \wh{H}_{N-1}\}$. Paralleling the arguments in the preceding subsections, her initial belief update is a function of $q_N$ as well as $\{\wh{H}_1, \ldots, \wh{H}_{N-1}\}$, but not of $\{q_1, \ldots, q_{N-1}\}$. Generalizing \eqref{eqn:Chuck_update_eq}, we have
\begin{equation} \label{eqn:Norah_update}
\begin{aligned}
	\frac{q_N^{\wh{h}_1, \ldots \wh{h}_{N-1}}}{1-q_N^{\wh{h}_1, \ldots \wh{h}_{N-1}}} &= \frac{q_N}{1-q_N} \frac{p_{\wh{H}_1|H} (\wh{h}_1|0)_{[N]}}{p_{\wh{H}_1|H} (\wh{h}_1|1)_{[N]}} \times \\
	&\prod_{n=2}^{N-1} \frac{p_{\wh{H}_n|\wh{H}_{n-1}, \ldots, \wh{H}_1,H}(\wh{h}_n|\wh{h}_{n-1}, \ldots, \wh{h}_1, 0)_{[N]}}{p_{\wh{H}_n|\wh{H}_{n-1}, \ldots, \wh{H}_1,H}(\wh{h}_n|\wh{h}_{n-1}, \ldots, \wh{h}_1, 1)_{[N]}}.
\end{aligned}
\end{equation}

Combining all observations, we obtain the following theorem. Define the initial belief update function for $N$th agent, $U_N$ as
\begin{align*}
q_N^{\wh{h}_1 \ldots \wh{h}_{N-1}} = U_N(q_N, \wh{h}_1, \wh{h}_2, \ldots, \wh{h}_{N-1}; N).
\end{align*}
\begin{theorem}
	The function $U_n, n \le N$ yielding the posterior belief of $N$th agent has the following recurrence relation:
	\begin{itemize}
		\item For $n = 1$, $U_1(q ; N) = q$.
		
		\item For $n > 1$,
		\begin{subequations} \label{eqn:general_update}
		\begin{align}
		&U_n(q, \wh{h}_1, \ldots, \wh{h}_{n-2}, 0; N) \nonumber \\
		&~~~ ~~~ = \frac{\tilde{q}}{\tilde{q} + (1-\tilde{q}) \frac{\int_{-\infty}^{\lambda_N(\tilde{q})} f_{Y_{N}|H}(y|1)dy }{\int_{-\infty}^{\lambda_N(\tilde{q})} f_{Y_{N}|H}(y|0)dy} }, \\
		& U_n(q, N, \wh{h}_1, \ldots, \wh{h}_{n-2}, 1; N) \nonumber \\
		&~~~ ~~~ = \frac{\tilde{q}}{\tilde{q} + (1-\tilde{q}) \frac{\int_{\lambda_N(\tilde{q})}^{\infty} f_{Y_{N}|H}(y|1)dy }{\int_{\lambda_N(\tilde{q})}^{\infty} f_{Y_{N}|H}(y|0)dy} },
		\end{align}
		\end{subequations}
		where $\tilde{q} = U_{n-1} (q, \wh{h}_1, \ldots, \wh{h}_{n-2}; N)$.
	\end{itemize}
\end{theorem}
Note that capital $N$ in (17a) and (17b) indicate the recursive updates are computed from the value that the $N$th agent thinks.

\begin{figure}[t]
	\centering
	\begin{subfigure}[t]{0.5\textwidth}
		\centering
		\includegraphics[width=3.5in]{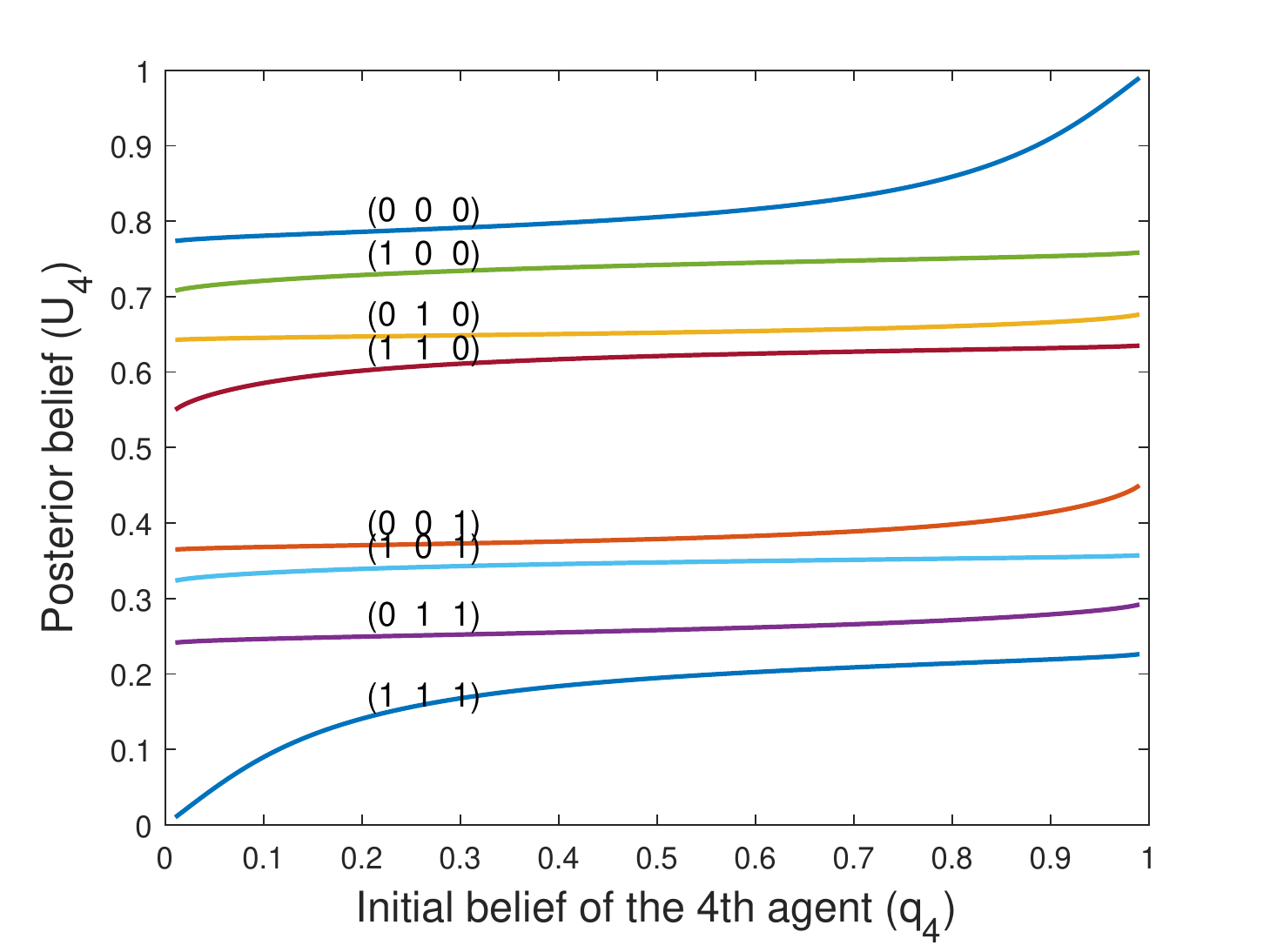}
		\subcaption{Noise variance $\sigma^2 = 1$}
	\end{subfigure}

	\begin{subfigure}[t]{0.5\textwidth}
		\centering
		\includegraphics[width=3.5in]{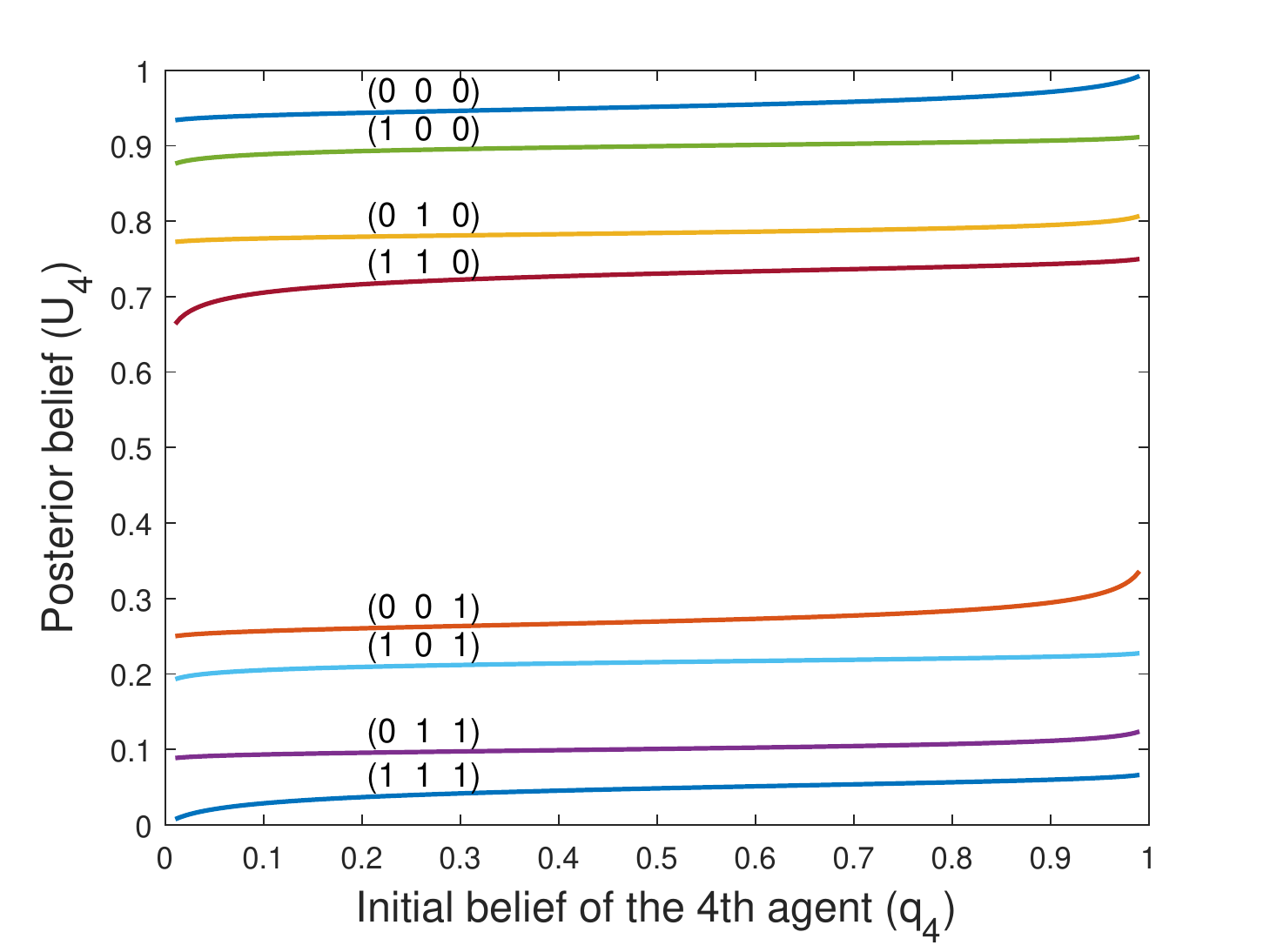}
		\subcaption{Noise variance $\sigma^2 = 0.25$}
	\end{subfigure}
	\caption{The function $U_4(q_4, \wh{h}_1, \wh{h}_2, \wh{h}_3;4)$---posterior belief of the fourth agent ($q_4^{\wh{h}_1, \wh{h}_2, \wh{h}_3}$)---for each possible combination of Alexis's, Blake's, and Chuck's decisions $[\wh{h}_1,\wh{h}_2,\wh{h}_3]$ when $c_{10} = c_{01} = 1$ and private signals are distorted by additive Gaussian noise with two noise levels. The posterior belief is mostly dependent on Chuck's decision; the top four curves are for $\wh{h}_3 = 0$ and the bottom four curves are for $\wh{h}_3 = 1$.}
	\label{fig:belief_update}
\end{figure}

Fig.~\ref{fig:belief_update} depicts the function $U_4(q_4, \wh{h}_1, \wh{h}_2, \wh{h}_3; 4)$ for $N = 4$ for eight possible combinations of Alexis's, Blake's, and Chuck's decisions $(\wh{h}_1, \wh{h}_2, \wh{h}_3)$. An interesting property of $U_N$ is that the posterior belief is much more dependent on the most recent decision $\wh{h}_{N-1}$ than on the earlier decisions $(\wh{h}_1, \ldots, \wh{h}_{N-2})$. This is because later acting agents consider more previous decisions, and hence more information than the first agents, their decisions should carry more weight. In this sense, we can say that recent decisions give more information than earlier decisions. This is especially the case when the $(N-1)$th agent has not followed precedent. This is because the $N$th agent rationally concludes that the $(N-1)$th agent observed strong evidence to justify a deviation from precedent. For example, if the decision history of the first five agents is $(0,0,0,0,1)$ then the sixth agent takes the last decision $1$ seriously even though the first four agents chose $0$. A reversal of an arbitrarily long precedent sequence may occur because we assume unbounded private signals; if private signals are bounded \cite{Banerjee1992,BikhchandaniHW1998}, then the influence of the precedent can reach a point where agents cannot receive a signal strong enough to justify a decision running counter to precedent. Another interesting point is that smaller noise variance changes beliefs more. It is clear from \eqref{eqn:general_update}, but also reasonable that when the variance is smaller, the $N$th agent trusts and is more inclined towards previous decisions. Note even though the prior updates of Norah in Fig.~\ref{fig:belief_update} do not depend on $\{q_1, \ldots, q_{N-1}\}$ and their corresponding likelihoods, the probability of prior decisions depends on them and implicitly, so does Norah's decision.

As we can see in Fig.~\ref{fig:belief_update}, the dominant previous decision for agent $N$ is the decision of agent $(N-1)$. We can prove that observing the $(N-1)$th agent's decision $0$ (or decision $1$), the $N$th agent's posterior belief becomes larger (or smaller), which in turn implies that the decision threshold of $N$th agent becomes larger (or smaller) so that she is more likely to declare decision $0$ (or $1$) as well.

\begin{theorem} \label{thm:order_preserve1}
	Suppose that noises are independent and additive, and have continuous densities. Fix some prior decisions $\{\wh{h}_1,\dots,\wh{h}_{N-2}\}$ and let $\tilde{q}_N, \tilde{q}_N^0, \tilde{q}_N^1$ denote the posterior beliefs of the $N$th agent given the $(N-2)$ decisions only, the $(N-2)$ decisions with $\wh{h}_{N-1} = 0$, and the $(N-2)$ decisions with $\wh{h}_{N-1} = 1$. Then,
	\begin{align*}
		\tilde{q}_N^1 < \tilde{q}_N < \tilde{q}_N^0.
	\end{align*}
\end{theorem}
\begin{IEEEproof}
	We know that $\tilde{q}_N, \tilde{q}_N^0, \tilde{q}_N^1$ differ only by the last multiplicative term of \eqref{eqn:Norah_update}. Since $\frac{q}{1-q}$ is monotone increasing, the statement is equivalent to showing:
	\begin{align*}
		\frac{\int_{\lambda_N(\tilde{q}_N)}^{\infty} f_{Y_N|H}(y|0)dy}{\int_{\lambda_N(\tilde{q}_N)}^{\infty} f_{Y_N|H}(y|1)dy} < 1 < \frac{\int_{-\infty}^{\lambda_N(\tilde{q}_N)} f_{Y_N|H}(y|0)dy}{\int_{-\infty}^{\lambda_N(\tilde{q}_N)} f_{Y_N|H}(y|1)dy}.
	\end{align*}
	Since the noise is independent and additive, $f_{Y_N|H}(y|1) = f_{Y_N|H}(y-1|0)$ so the term on the left side
	\begin{align*}
		&\frac{\int_{\lambda_N(\tilde{q}_N)}^{\infty} f_{Y_N|H}(y|0)dy}{\int_{\lambda_N(\tilde{q}_N)}^{\infty} f_{Y_N|H}(y|1)dy} = \frac{\int_{\lambda_n(\tilde{q}_N)}^{\infty} f_{Y_N|H}(y|0)dy}{\int_{\lambda_N(\tilde{q}_N)-1}^{\infty} f_{Y_N|H}(y|0)dy} \\
		&= \frac{\int_{\lambda_N(\tilde{q}_N)}^{\infty} f_{Y_N|H}(y|0)dy}{\int_{\lambda_N(\tilde{q}_N)-1}^{\lambda_N(\tilde{q}_N)} f_{Y_N|H}(y|0)dy + \int_{\lambda_N(\tilde{q}_N)}^{\infty} f_{Y_N|H}(y|0)dy} < 1.
	\end{align*}
	The right inequality can be shown similarly.
\end{IEEEproof}

Considering the complicated relationships that individual decisions have on the evolution of initial beliefs, it is also important to verify if the belief evolution preserves the ordering, given the same set of subsequent decisions. That is, given two beliefs $q_L < q_R$ at some point of the recursive update and the same sequence of following $d$ decisions, then it is important to characterize the likelihoods for which the the ordering is preserved in the resulting posterior beliefs, given the sequence of decisions, which is described in the following theorem.

\begin{theorem} \label{thm:order_preserve2}
	Suppose that noise is independent and additive, and has a continuous density. Consider two beliefs $q_L < q_R$. Then, for any given later-acting decisions $d$, the posterior belief satisfies $q_L^{d} < q_R^{d}$ if and only if
	\begin{align}
		g_1(q) &:= \frac{q}{1-q} \frac{\int_{-\infty}^{\lambda_N(q)} f_{Y_N|H}(y|0)dy}{\int_{-\infty}^{\lambda_N(q)} f_{Y_N|H}(y|1)dy}, \label{eqn:order_preserve1}\\
		g_2(q) &:= \frac{q}{1-q} \frac{\int_{\lambda_N(q)}^{\infty} f_{Y_N|H}(y|0)dy}{\int_{\lambda_N(q)}^{\infty} f_{Y_N|H}(y|1)dy} \label{eqn:order_preserve2}
	\end{align}
	are both increasing in $q$.
\end{theorem}
\begin{IEEEproof}
	Note that once observing decision $0$, beliefs are updated as
	\begin{align*}
		\frac{q_L^0}{1-q_L^0} &= \frac{q_L}{1-q_L} \frac{\int_{-\infty}^{\lambda_N(q_L)} f_{Y_N|H}(y|0)dy}{\int_{-\infty}^{\lambda_N(q_L)} f_{Y_N|H}(y|1)dy}, \\
		\frac{q_R^0}{1-q_R^0} &= \frac{q_R}{1-q_R} \frac{\int_{-\infty}^{\lambda_N(q_R)} f_{Y_N|H}(y|0)dy}{\int_{-\infty}^{\lambda_N(q_R)} f_{Y_N|H}(y|1)dy},
	\end{align*}
	and so if \eqref{eqn:order_preserve1} holds, $q_L^0 < q_R^0$. Similarly, \eqref{eqn:order_preserve2} can be shown by updating after decision $1$.
\end{IEEEproof}

Let us state some properties of Mills ratio \cite{Mills1926, Sampford1953}, which is about Gaussian distribution, and we will see that $g_1(q), g_2(q)$ are both increasing if likelihood is Gaussian.

\begin{lemma}[\cite{Sampford1953}] \label{lem:Mills_ratio}
	Define $\eta(x) := \phi(x) / Q(x)$, the inverse of Mills ratio. Then, for any $x \in \mathbb{R}$, it is true that $0 < \eta'(x) < 1$ and $\eta''(x) > 0$.
\end{lemma}

\begin{corollary} \label{lem:inc_fn_gauss}
Consider a Gaussian likelihood, i.e., $Y_N = H + Z_N$, where $Z_N$ are independent and identically drawn from $\calN(0, \sigma^2)$, for some $\sigma^2 > 0$. Then $g_1(q), g_2(q)$ are both increasing in $q$.
\end{corollary}
\begin{IEEEproof}
Let us consider $g_2(q)$ first. For the binary hypothesis test with Gaussian noise, we know that the decision threshold for the likelihood ratio test is given by
\[
\lambda_N(q) = \frac{1}{2} + \sigma^2 \log \pth{\frac{c_{10}q}{c_{01}(1-q)}}.
\]
Then, we have
\[
g_2(q) = \frac{q}{1-q} \frac{Q\pth{\tfrac{\lambda_N(q)}{\sigma}}}{Q\pth{\tfrac{\lambda_N(q)-1}{\sigma}}}.
\]
Letting $x := \log \tfrac{c_{10}q}{c_{01}(1-q)}$, it is sufficient to show that
\begin{align*}
\tilde{g}(x) &:= \log \pth{ \frac{c_{10}}{c_{01}} g_2(q) } \\
&= x + \log \pth{Q\pth{\sigma x + \tfrac{1}{2\sigma}}} - \log \pth{Q\pth{\sigma x - \tfrac{1}{2\sigma}}},
\end{align*}
is increasing in $x$ since $c_{10},c_{01}$ are positive constants, $\log (\cdot)$ is a monotonically increasing function, and $x$ is a strictly increasing function of $q$.

The first derivative of $\tilde{g}$ is given by
\begin{equation} \label{eqn:gtilde_der}
\tilde{g}'(x) = 1 - \sigma\eta\pth{\sigma x + \tfrac{1}{2\sigma}} + \sigma\eta\pth{\sigma x - \tfrac{1}{2\sigma}}.
\end{equation}
Since $\eta(\cdot)$ is a continuous function, using the mean value theorem, there exists $y \in \pth{\sigma x - \tfrac{1}{2\sigma},\sigma x + \tfrac{1}{2\sigma}}$, such that
\begin{equation} \label{eqn:eta_mvt_exp}
\sigma\eta\pth{\sigma x + \tfrac{1}{2\sigma}} - \sigma\eta\pth{\sigma x - \tfrac{1}{2\sigma}} = \sigma \eta'(y)\frac{1}{\sigma} = \eta'(y).
\end{equation}
From the first property of Lem.~\ref{lem:Mills_ratio}, $0 < \eta'(y) < 1$, we have
\[
\eta\pth{\sigma x + \tfrac{1}{2\sigma}} - \eta\pth{\sigma x - \tfrac{1}{2\sigma}} < 1.
\]
Thus, from \eqref{eqn:gtilde_der}, it follows that $\tilde{g}'(x) > 0$ for all $x$, indicating that $\tilde{g}(\cdot)$ is an increasing function of $x$. This in turn implies that $g_2(\cdot)$ is also an increasing function.

To prove the result for $g_1$, it is sufficient to observe that by the symmetry of error probabilities:
\[
g_1(q) = \frac{1}{g_2(1-q)}.
\]
\end{IEEEproof}

\section{Optimal Belief} \label{sec:optimal_belief}
We described the initial belief evolution and decision-making model in Sec.~\ref{sec:belief_update}. In this section, we investigate the set of initial beliefs that minimize the Bayes risk. We consider the case of two agents for analytical tractability although the broad nature of the arguments extend to multi-agent systems. Note that the Bayes risk of the system with $N=2$ is the same as Blake's Bayes risk because his decision is adopted as the final decision.

Let us recapitulate the computation of Blake's Bayes risk. Alexis chooses her decision threshold as $\lambda_1 := \lambda_1(q_1)$. Her probabilities of error are given by
\begin{align*}
P_{e,1}^{\textrm{I}} &= p_{\wh{H}_1|H}(1|0) = \int_{\lambda_1}^{\infty} f_{Y_1|H}(y|0) dy, \\
P_{e,1}^{\textrm{II}} &= p_{\wh{H}_1|H}(0|1) = \int_{-\infty}^{\lambda_1} f_{Y_1|H}(y|1) dy.
\end{align*}

Blake however presumes Alexis uses the decision threshold $\lambda_{1, [2]} := \lambda_2(q_2)$ and computes her probabilities of error accordingly\footnote{Recall that the subscript $[2]$ denotes the quantity `seen by' Blake.}:
\begin{align*}
P_{e,1,[2]}^{\textrm{I}} &= p_{\wh{H}_1|H}(1|0)_{[2]} = \int_{\lambda_{1,[2]}}^{\infty} f_{Y_2|H}(y|0) dy, \\
P_{e,1,[2]}^{\textrm{II}} &= p_{\wh{H}_1|H}(0|1)_{[2]} = \int_{-\infty}^{\lambda_{1,[2]}} f_{Y_2|H}(y|1) dy.
\end{align*}

When Alexis decides $\wh{H}_1 = 0$, Blake updates his belief $q_2$ to the posterior $q_2^0$:
\begin{equation} \label{eqn:q20_posterior}
\begin{aligned}
&\frac{q_2^0}{1-q_2^0} = \frac{q_2}{1-q_2} \frac{1-P_{e, 1, [2]}^{\textrm{I}}}{P_{e, 1, [2]}^{\textrm{II}}} \\
&\implies q_2^0 = \frac{q_2(1-P_{e,1,[2]}^{\textrm{I}})}{q_2(1-P_{e, 1, [2]}^{\textrm{I}}) + (1-q_2) P_{e, 1, [2]}^{\textrm{II}}},
\end{aligned}
\end{equation}
his decision threshold is $\lambda_2^0 := \lambda_2(q_2^0)$, and the probabilities of error are
\begin{align*}
P_{e,2}^{\textrm{I}_0} &= p_{\wh{H}_2|\wh{H}_1, H}(1|0,0) = \int_{\lambda_2^0}^{\infty} f_{Y_2|H}(y|0) dy, \\
P_{e,2}^{\textrm{II}_0} &= p_{\wh{H}_2|\wh{H}_1, H}(0|0,1) = \int_{-\infty}^{\lambda_2^0} f_{Y_2|H}(y|1) dy.
\end{align*}

Likewise, when Alexis decides $\wh{H}_1 = 1$, Blake updates his belief $q_2$ to the posterior $q_2^1$:
\begin{equation} \label{eqn:q21_posterior}
\begin{aligned}
&\frac{q_2^1}{1-q_2^1} = \frac{q_2}{1-q_2} \frac{P_{e,1,[2] }^{\textrm{I}}}{1-P_{e,1,[2]}^{\textrm{II}}} \\
&\implies q_2^1 = \frac{q_2 P_{e, 1,[2]}^{\textrm{I}}}{q_2 P_{e,1,[2]}^{\textrm{I}} + (1-q_2) (1-P_{e, 1,[2]}^{\textrm{II}})}, 
\end{aligned}
\end{equation}
his decision threshold is $\lambda_2^1 := \lambda_2(q_2^1)$, and the probabilities of error are
\begin{align*}
P_{e,2}^{\textrm{I}_1} &= p_{\wh{H}_2|\wh{H}_1, H}(1|1,0) = \int_{\lambda_2^1}^{\infty} f_{Y_2|H}(y|0) dy, \\
P_{e,2}^{\textrm{II}_1} &= p_{\wh{H}_2|\wh{H}_1, H}(0|1,1) = \int_{-\infty}^{\lambda_2^1} f_{Y_2|H}(y|1) dy.
\end{align*}

Now we compute the system's Bayes risk (or Blake's Bayes risk) $R_2$:
\begin{align}
R_2 &= c_{10} p_{\wh{H}_2,H}(1,0) + c_{01}p_{\wh{H}_2, H}(0,1) \nonumber \\
&= c_{10} \sum_{\wh{h}_1 \in \{0,1\} } p_{\wh{H}_2|\wh{H}_1,H}(1|\wh{h}_1,0) p_{\wh{H}_1|H}(\wh{h}_1|0) p_H(0)  \nonumber \\
&~~~ + c_{01} \sum_{\wh{h}_1 \in \{0,1\} } p_{\wh{H}_2|\wh{H}_1,H}(0|\wh{h}_1,1) p_{\wh{H}_1|H}(\wh{h}_1|1) p_H(1) \nonumber \\
&= c_{10}\left[ P_{e,2}^{\textrm{I}_0} (1-P_{e,1}^{\textrm{I}}) + P_{e,2}^{\textrm{I}_1} P_{e,1}^{\textrm{I}} \right] p_0 \nonumber \\
&~~~ + c_{01}\left[ P_{e,2}^{\textrm{II}_0} P_{e,1}^{\textrm{II}} + P_{e,2}^{\textrm{II}_1} (1-P_{e,1}^{\textrm{II}}) \right] (1-p_0). \label{eqn:Blake_Bayes_risk}
\end{align}

Note that the Bayes risk $R_2$ in \eqref{eqn:Blake_Bayes_risk} is a function of $q_1$ and $q_2$. One might think that $R_2$ is minimum at $q_1 = q_2 = p_0$ as Alexis makes the best decision for the true prior and Blake does not misunderstand her decision. Surprisingly, however, this turns out to not be true. We prove this by studying Alexis's optimal belief $q_1^*$ that minimizes $R_2$. 

\begin{theorem} \label{thm:optimal_beliefs}
	Alexis's and Blake's optimal beliefs $q_1^*, q_2^*$ that minimize $R_2$ satisfy
	\begin{align}
	\frac{q_1^*}{1-q_1^*} = \frac{p_0 (P_{e,2}^{\textrm{I}_1} - P_{e,2}^{\textrm{I}_0})}{(1-p_0) (P_{e,2}^{\textrm{II}_0} - P_{e,2}^{\textrm{II}_1})}. \label{eqn:optimal_belief}
	\end{align}
\end{theorem}

Before proceeding to the proof, note that error probability terms in the right-side are dependent on $q_2$, but not on $q_1$. Furthermore, the value of $(P_{e,2}^{\textrm{I}_1} - P_{e,2}^{\textrm{I}_0})/(P_{e,2}^{\textrm{II}_0} - P_{e,2}^{\textrm{II}_1})$ is generally not $1$, i.e., in general $q_1 = q_2 = p_0$ is not the optimal belief. For example, for the additive Gaussian noise model considered in the next section, the ratio is not equal to $1$ except when $p_0 = c_{01} / (c_{10} + c_{01})$.

\begin{IEEEproof}[Proof of Thm.~\ref{thm:optimal_beliefs}]
Let us consider the first derivative of \eqref{eqn:Blake_Bayes_risk} with respect to $q_1$:
\begin{align*}
	\frac{\partial R_2}{\partial q_1} &= c_{10} p_0 (P_{e,2}^{\textrm{I}_1} - P_{e,2}^{\textrm{I}_0}) \frac{\partial P_{e,1}^\textrm{I}}{\partial q_1} \\
	&~~~ + c_{01} (1-p_0) (P_{e,2}^{\textrm{II}_0} - P_{e,2}^{\textrm{II}_1}) \frac{\partial P_{e,1}^{\textrm{II}}}{\partial q_1}.
\end{align*}
We want to find $q_1$ that minimizes $R_2$, i.e., $q_1$ makes the first derivative zero. Using
\begin{align*}
	\frac{dP_{e,1}^{\textrm{I}}}{dq_1} &= \frac{dP_{e,1}^{\textrm{I}}}{d \lambda_1} \frac{d \lambda_1}{dq_1} =-f_{Y_1|H}(\lambda_1|0) \frac{d \lambda_1}{dq_1}, \\
	\frac{dP_{e,1}^{\textrm{II}}}{dq_1} &= \frac{dP_{e,1}^{\textrm{II}}}{d \lambda_1} \frac{d \lambda_1}{dq_1} = f_{Y_1|H}(\lambda_1|1) \frac{d \lambda_1}{dq_1};
\end{align*}
this occurs when
\begin{align}
	&c_{10}p_0 (P_{e,2}^{\textrm{I}_1} - P_{e,2}^{\textrm{I}_0}) f_{Y_1|H} (\lambda_1|0) \nonumber \\
	&= c_{01}(1-p_0) (P_{e,2}^{\textrm{II}_0} - P_{e,2}^{\textrm{II}_1}) f_{Y_1|H}(\lambda_1|1) \nonumber \\
	&\iff \frac{f_{Y_1|H}(\lambda_1|1)}{f_{Y_1|H}(\lambda_1|0)} = \frac{c_{10}p_0 (P_{e,2}^{\textrm{I}_1} - P_{e,2}^{\textrm{I}_0})}{c_{01}(1-p_0) (P_{e,2}^{\textrm{II}_0} - P_{e,2}^{\textrm{II}_1})}. \label{eqn:temp1}
\end{align}
Note that $\lambda_1 = \lambda_1(q_1)$ is the solution to \eqref{eqn:decision_threshold}, 
\begin{align}
	\frac{f_{Y_1|H}(\lambda_1|1)}{f_{Y_1|H}(\lambda_1|0)} = \frac{c_{10}q_1}{c_{01}(1-q_1)}. \label{eqn:temp2}
\end{align}
Equating \eqref{eqn:temp1} and \eqref{eqn:temp2} completes the proof.
\end{IEEEproof}

The theorem considers general continuous likelihoods $\{f_{Y_n|H}\}$ with the monotonicity assumption on $\lambda(q)$. It is interesting to evaluate the optimal beliefs in the case of Gaussian likelihoods (i.e., additive Gaussian noise) and obtain insights into optimality in the sequential decision-making problem.

\section{Gaussian Likelihoods} \label{sec:Gaussian_diverse_expertise}
\begin{figure}[t]
	\centering
	\includegraphics[width=3.5in]{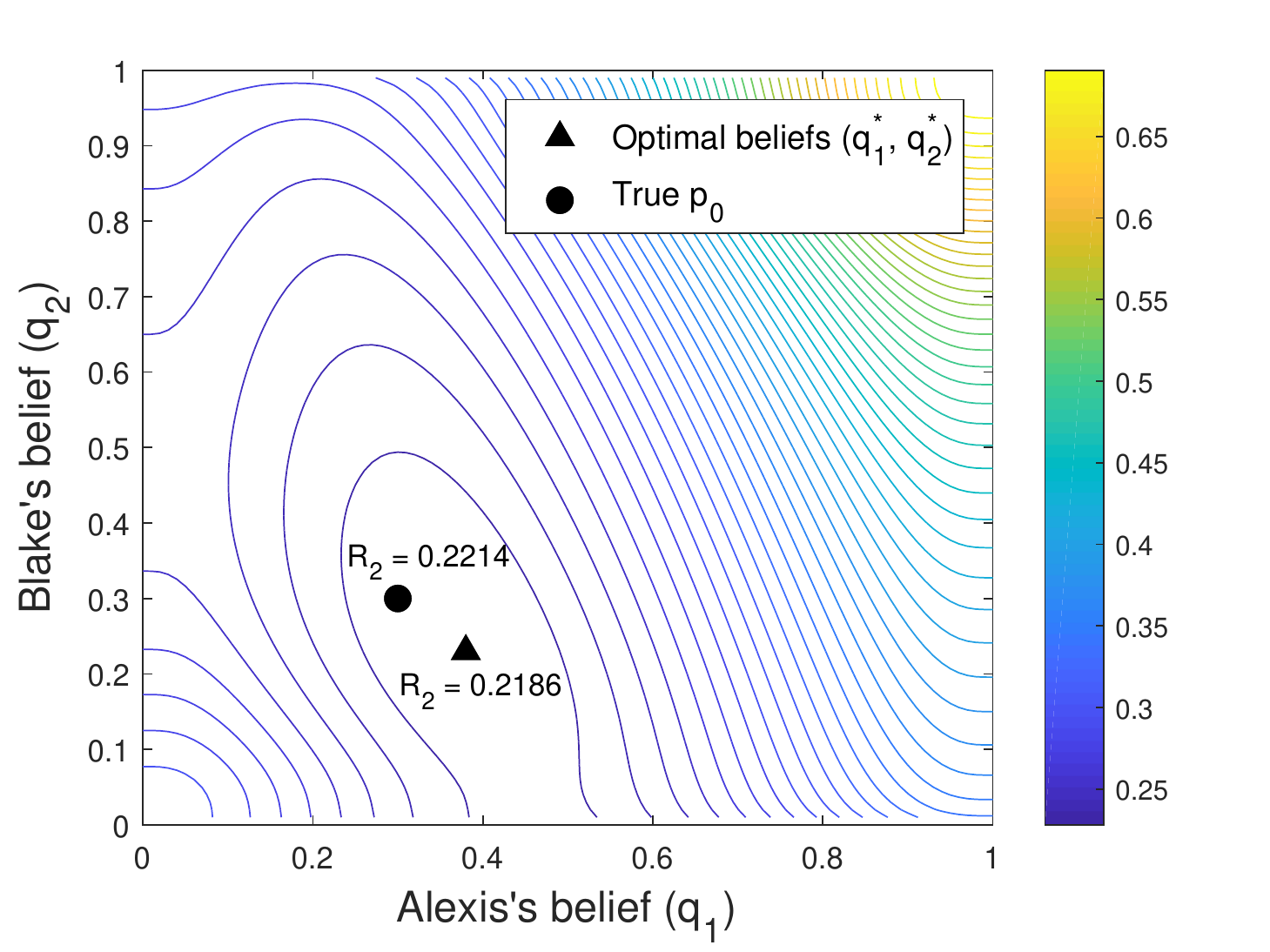}
	\caption{The Bayes risk for $q_1, q_2 \in (0,1)$ with $p_0 = 0.3$, $c_{10} = c_{01} = 1$, and additive standard Gaussian noise. The pair of optimal beliefs ($\blacktriangle$) yields $R_2=0.2186$, while the true prior ($\bullet$) yields $R_2 = 0.2214$.}
	\label{fig:Blake_Bayes_risk}
\end{figure}
We now focus on Gaussian likelihoods and study their optimal beliefs in this section. Suppose the $n$th agent receives the signal $Y_n = H + Z_n$, where $Z_n$ is an independent additive Gaussian noise with zero mean and variance $\sigma_n^2 > 0$. Thus, the received signal probability densities for $H = h$ are
\begin{align*}
f_{Y_n | H}(y_n | h)=\phi(y_n; h, \sigma_n^2).
\end{align*}
For a belief $q_n$, the decision threshold is then determined by the likelihood ratio test,
\begin{align*}
	\mathcal{L}_n(y_n) = \frac{f_{Y_n|H}(y_n|1)}{f_{Y_n|H}(y_n|0)} \underset{\wh{H}_1 = 0}{\overset{\wh{H}_1 = 1}{\gtrless}} \frac{c_{10}q_n}{c_{01}(1-q_n)},
\end{align*}
that simplifies to the following simple threshold condition for Gaussian likelihoods:
\begin{align}
	y_n \underset{\wh{H}_1 = 0}{\overset{\wh{H}_1 = 1}{\gtrless}} \lambda_n(q_n) = \frac{1}{2} + \sigma_n^2 \log \left( \frac{c_{10}q_n}{c_{01}(1-q_n)} \right). \label{eqn:Gauss_BHT_threshold}
\end{align}
Here the index $n$ represents the $n$th agent in the system, as the belief and variance of the agent varies along the chain.

\begin{figure}[t]
	\centering
	\begin{subfigure}[b]{0.24\textwidth}
		\includegraphics[width=1.75in]{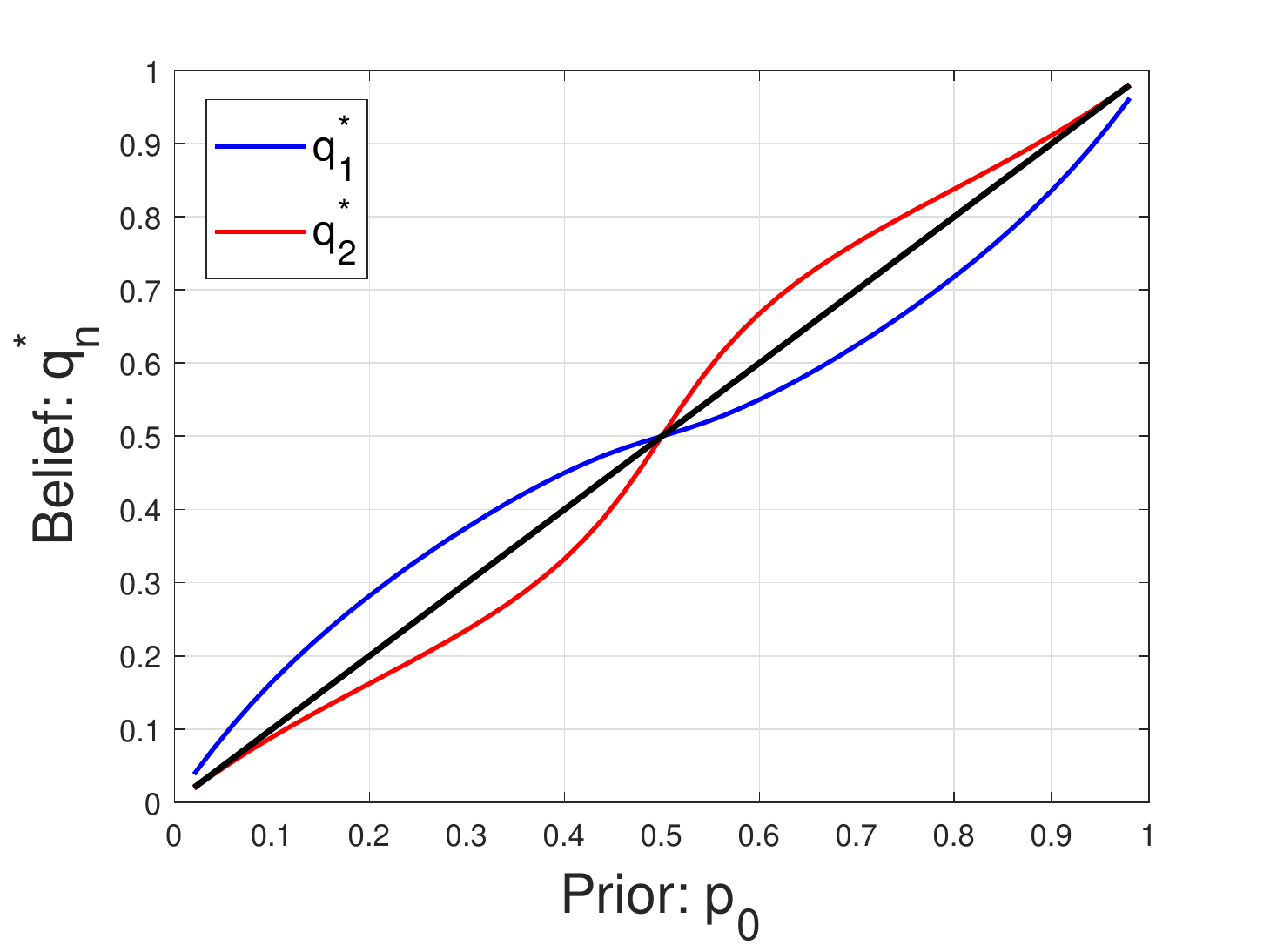}
		\subcaption{$c_{10} = c_{01} = 1$}
	\end{subfigure}
	\begin{subfigure}[b]{0.24\textwidth}
		\includegraphics[width=1.75in]{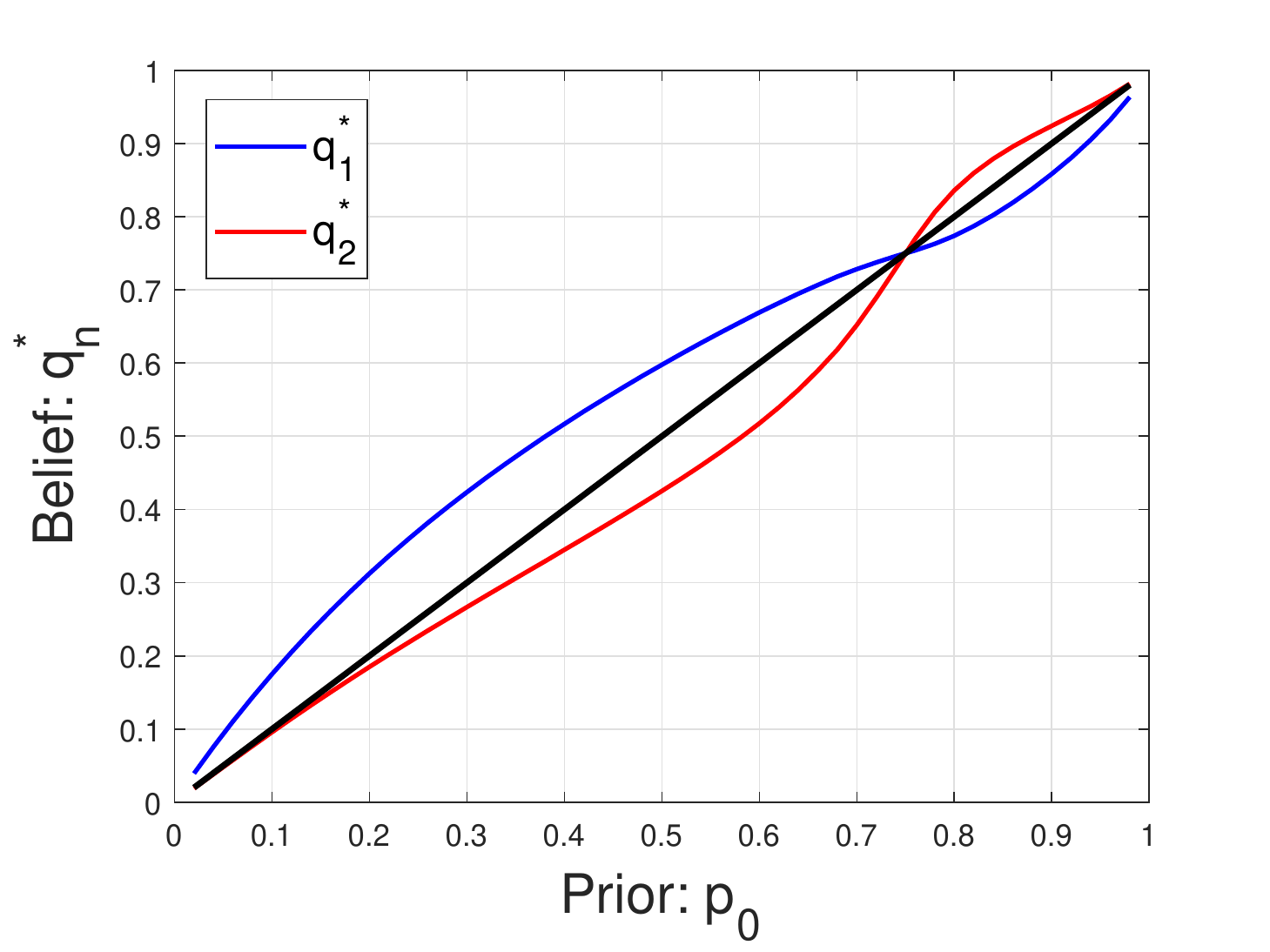}
		\subcaption{$c_{10} = 1, c_{01} = 3$}
	\end{subfigure}
	\caption{The trend of the optimal beliefs for $N=2$ (Alexis, Blake). $Z_1, Z_2$ are standard Gaussian.}
	\label{fig:2_agents_optimal_beliefs}
\end{figure}

\begin{figure}[t]
	\centering
	\begin{subfigure}[b]{0.24\textwidth}
		\includegraphics[width=1.75in]{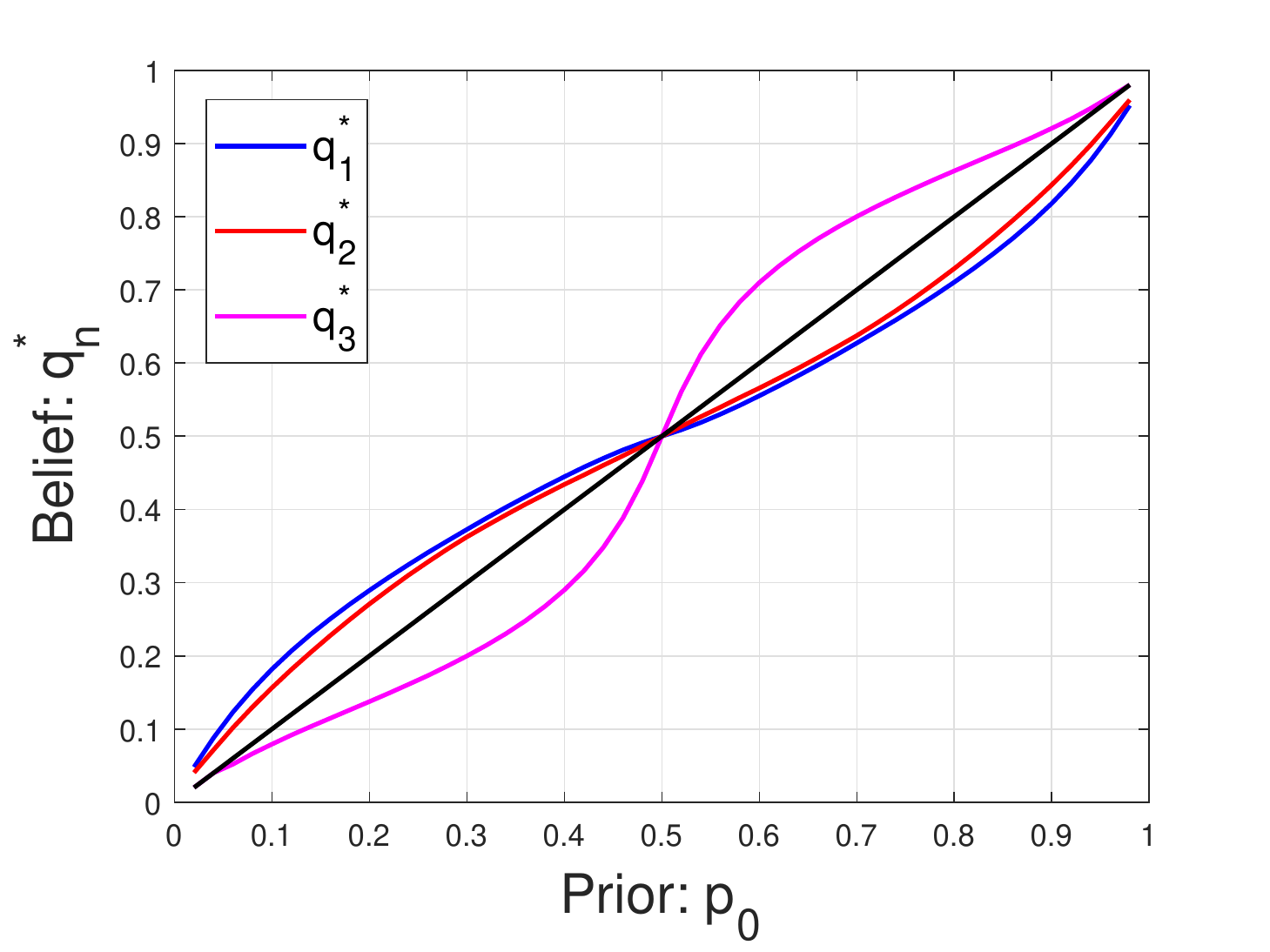}
		\subcaption{$c_{10} = c_{01} = 1$}
	\end{subfigure}	
	\begin{subfigure}[b]{0.24\textwidth}
		\includegraphics[width=1.75in]{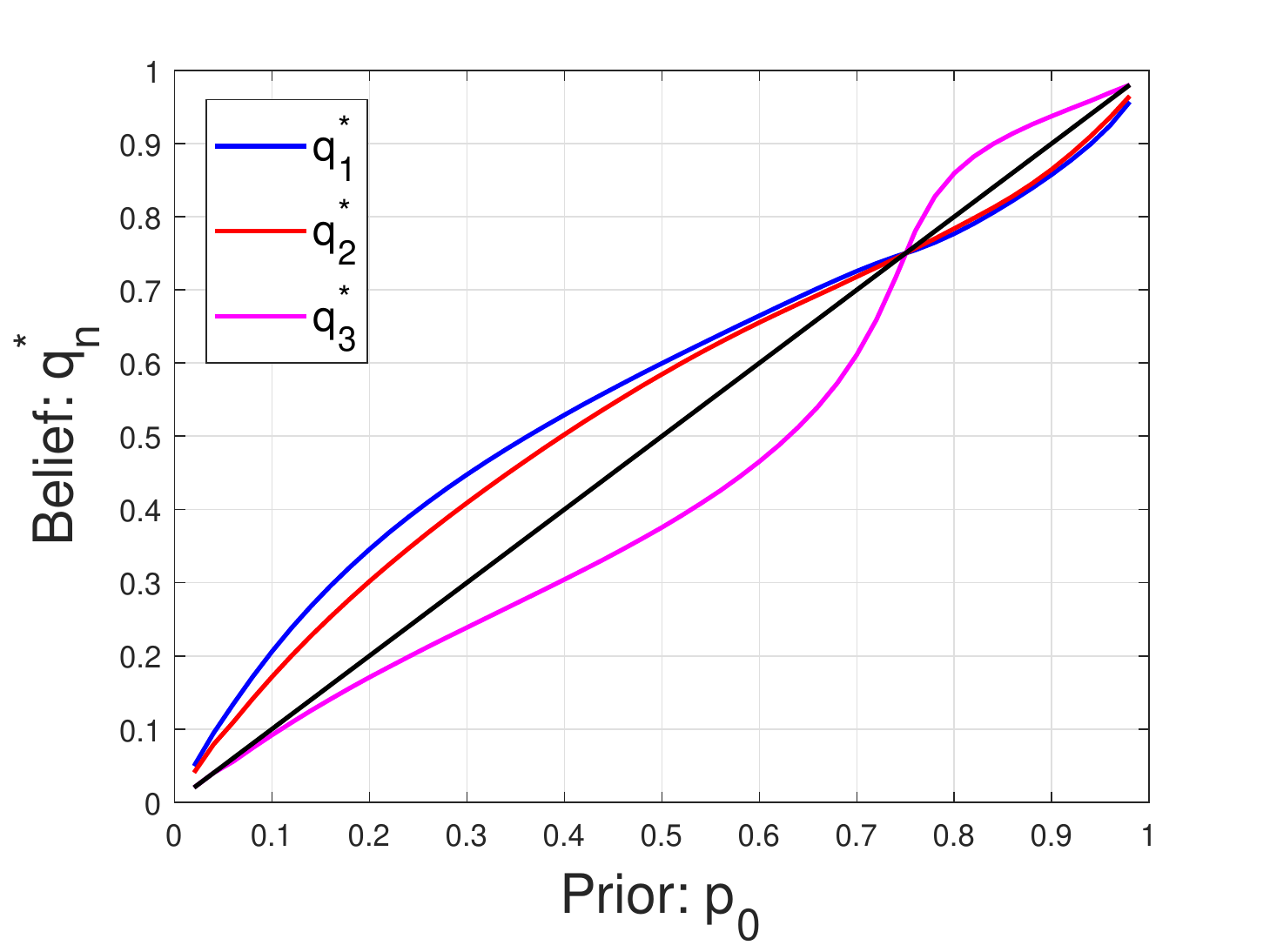}
		\subcaption{$c_{10} = 1, c_{01} = 3$}
	\end{subfigure}
	\caption{The trend of the optimal beliefs for $N=3$ (Alexis, Blake, and Chuck). $Z_1, Z_2, Z_3$ are standard Gaussian.}
	\label{fig:3_agents_optimal_beliefs}
\end{figure}

\begin{figure}[t]
	\centering
	\includegraphics[width=3.5in]{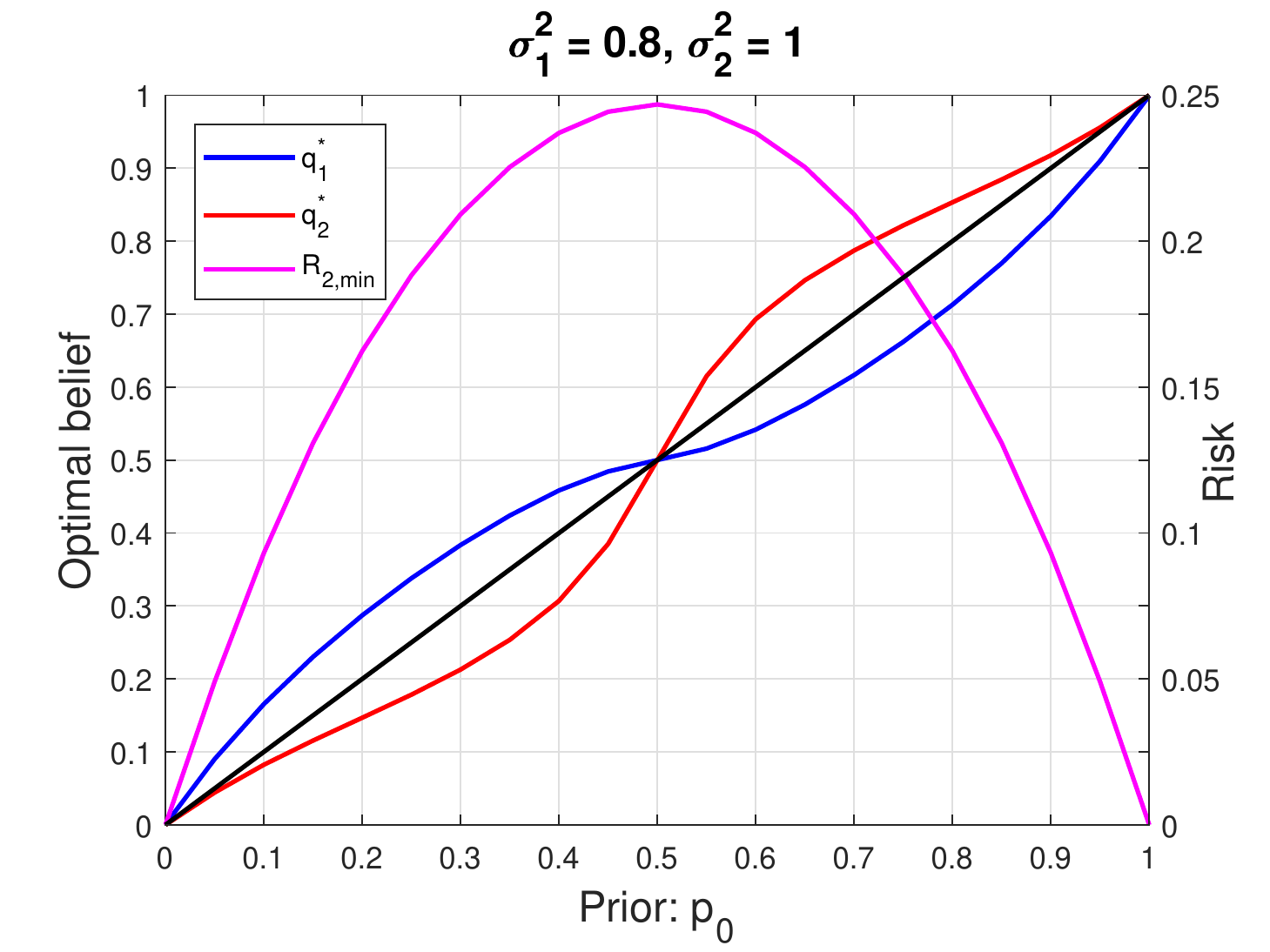}
	\caption{Optimal beliefs when the preceding agent has smaller noise, where $\sigma_1^2 = 0.8$ and $\sigma_2^2=1$.}
	\label{fig:sigma1<sigma2}
\end{figure}

\begin{figure}[t]
	\centering
	\includegraphics[width=3.5in]{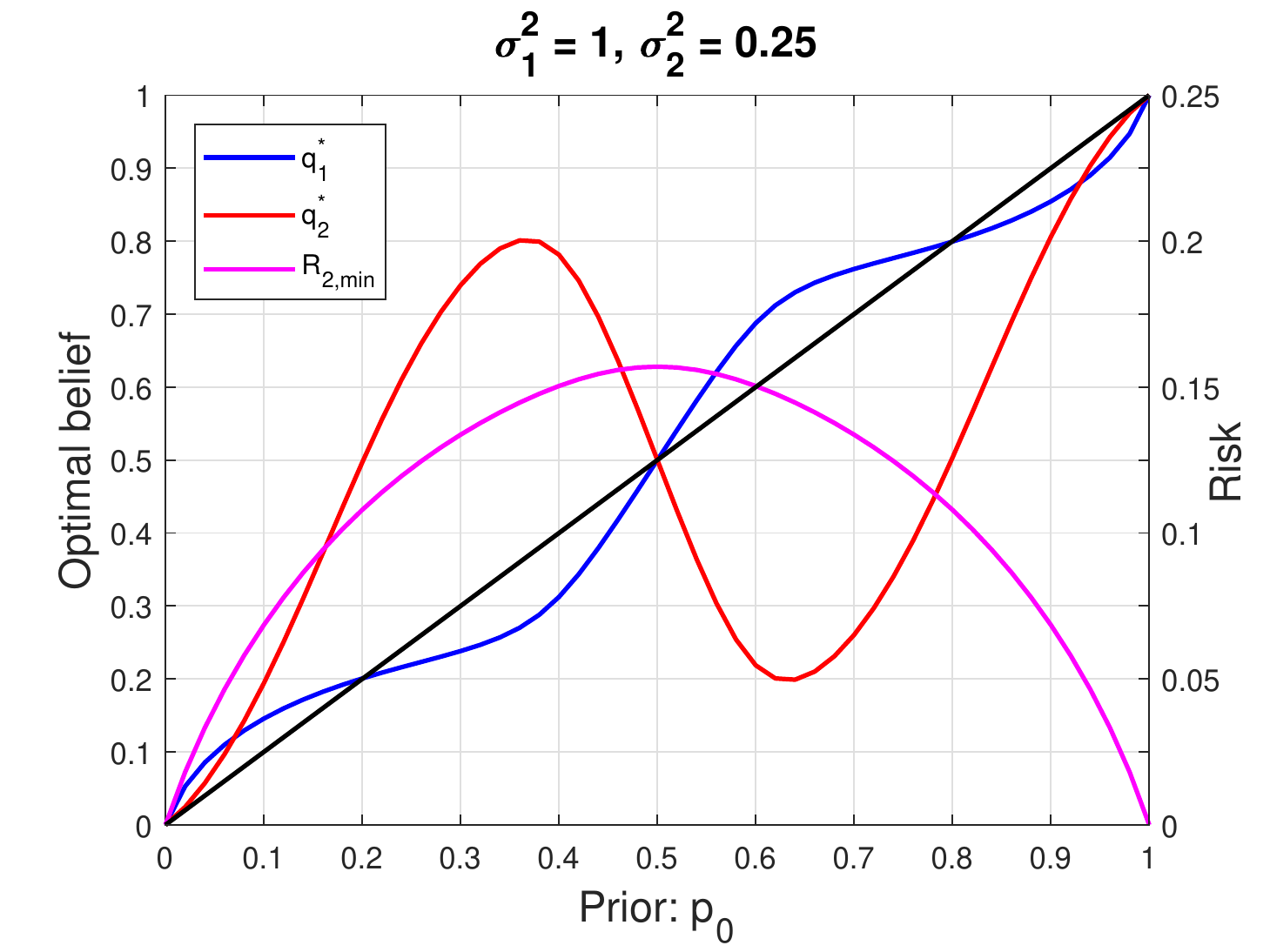}
	\caption{Optimal beliefs when the later-acting agent has smaller noise, where $\sigma_1^2 = 1$ and $\sigma_2^2=0.25$.}
	\label{fig:sigma1>sigma2}
\end{figure}

Using the recursive update in Sec.~\ref{sec:belief_update} and decision threshold \eqref{eqn:Gauss_BHT_threshold}, it is possible to obtain the Bayes risk of Blake (i.e., $N=2$) for given beliefs $q_1, q_2$. Fig.~\ref{fig:Blake_Bayes_risk} depicts Blake's Bayes risk for $q_1, q_2 \in (0,1)$, and explicitly shows that knowing true prior probability is not optimal. The social learning problem with Bayes costs $c_{10} = c_{01} = 1$, prior $p_0 = 0.3$, and additive Gaussian noise with zero mean and unit variance results in a Bayes risk that is minimum when Alexis's belief is $0.38$ and Blake's belief is $0.23$, shown in the figure (triangle) and is compared to the true prior (circle).

Figs.~\ref{fig:2_agents_optimal_beliefs} and \ref{fig:3_agents_optimal_beliefs} show the trend of optimal belief pair that minimizes the last agent's Bayes risk, when all agents have the same noise levels for the case of two and three agents respectively. We can observe several common characteristics. First, the non-terminal agents (i.e., Alexis for $N=2$ and Alexis and Blake for $N=3$) overweight their beliefs if $p_0$ is small and underweight it if $p_0$ is large. We call this \emph{open-minded} behavior as it enhances less likely events. Second, the last agent (i.e., Blake for $N=2$ and Chuck for $N=3$) underweights the belief if $p_0$ is small and overweights it if $p_0$ is large, implicitly compensating for the biases of the preceding agents. Such behavior is referred to as being \emph{closed-minded} as it represents a cautious outlook to the decision-making problem. Lastly, there is a unique, non-trivial prior, $p_0 \in (0,1)$, where all agents' optimal beliefs are identical to the true prior.

However, the case of nonidentical noise variances of agents results in a very different behavior of optimal beliefs, especially when the last agent has smaller noise. The optimal beliefs for $N=2$ and the case of the preceding agent having smaller noise, and that of the last agent having smaller noise respectively are shown in Figs.~\ref{fig:sigma1<sigma2} and \ref{fig:sigma1>sigma2}. As can be observed, the optimal belief curves are markedly different when the last agent has smaller noise, and we now derive some analytical properties of $q_1^*,q_2^*$.

\begin{theorem} \label{prop:belief_props}
For any $\sigma_1^2$ and $\sigma_2^2$, $q_1^*$ and $q_2^*$ satisfy:
\begin{enumerate}
\item for $p_0 \in (0,1)$, $q_1^* \le p_0$ if and only if $q_2^* \geq \frac{c_{01}}{c_{01}+ c_{10}}$, with equality for $q_2^* = \frac{c_{01}}{c_{01}+ c_{10}}$.
\item $p_0 = q_1^* = q_2^*$ if and only if $p_0 \in \sth{0,\frac{c_{01}}{c_{01}+ c_{10}},1}$.
\end{enumerate}
\end{theorem}
\begin{IEEEproof}
Given in App.~\ref{app:belief_props}.
\end{IEEEproof}

Thm.~\ref{prop:belief_props} highlights the fact that if the last agent believes the null hypothesis is more likely, then the ideal predecessor underweights the prior, and vice versa. Additionally, for $p_0$ near zero (near one) the optimal predecessor overweights (underweights) the prior.

In particular, let us consider two cases separately. First, let the predecessor have smaller noise. Then the curves for optimal beliefs and the corresponding Bayes risk are as shown in Fig.~\ref{fig:sigma1<sigma2}. The behavior here is similar to the case with equal noise, indicating that the reducing noise of the predecessor does not alter the overall behaviors of beliefs, as the last agent is unaware of this improved signal quality.

On the other hand, when the last agent has smaller noise, we notice that the nature of curves changes, as shown in Fig.~\ref{fig:sigma1>sigma2}. The behavior of the ideal agents indicates that when the predecessor has significantly larger noise than the last agent, the last agent stays open-minded. In addition, $q_1^*$ has multiple crossings with $p_0$, but $q_2^*$ has a single crossing at $q_2^* = c_{01}/(c_{01}+c_{10})$.

As expected, the ideal predecessor is open-minded for near-deterministic priors ($p_0$ close to zero or one). However, when the prior uncertainty in the hypotheses is high ($p_0$ near $1/2$), we note that the ideal last agent with less noise favors the less likely hypothesis. This can be attributed to the fact that the last agent stays open-minded to the less likely hypothesis when the predecessor with larger noise is more likely to make errors. To further understand the nature of such an predecessor, we characterize the crossings of the optimal belief curve with the prior $q_1^* = p_0$ .

\begin{theorem} \label{thm:fixed_pt}
	The set of all $p_0$ such that $q_1^* = p_0$, $q_2^* = \frac{c_{01}}{c_{01}+c_{10}}$ is given by the solutions to
	\begin{align} \label{eqn:fixed_pt_eqn}
	e^x = \frac{1 - \beta Q(-\alpha + \sigma_1 x)}{1 - \beta Q(-\alpha - \sigma_1 x)},
	\end{align}
	where
	\begin{align*}
	x = \log\pth{\frac{c_{10}p_0}{c_{01}(1-p_0)}}, ~ \alpha = \frac{1}{2\sigma_1}, ~ \beta = 1-\frac{Q\pth{1/2\sigma_2}}{Q\pth{-1/2\sigma_2}}.
	\end{align*}
\end{theorem}
\begin{IEEEproof}
	Given in App.~\ref{app:fixed_pt}.
\end{IEEEproof}
We note that $p^* = \frac{c_{01}}{c_{01} + c_{10}}$ is always a solution to \eqref{eqn:fixed_pt_eqn}. The case of multiple solutions to \eqref{eqn:fixed_pt_eqn} is of particular interest and a sufficient condition is given in the following corollary.
\begin{corollary} \label{cor:suff_cond}
	If 
	\begin{equation} \label{eqn:suff_cond}
	\frac{2\beta \sigma_1 \phi{(\alpha)}}{1-\beta Q(-\alpha)} > 1,
	\end{equation}
	then, \eqref{eqn:fixed_pt_eqn} has at least $3$ solutions in $(0,1)$.
\end{corollary}
\begin{IEEEproof}
	Since $x$ is a monotonic function of $p_0$, it is sufficient to show that \eqref{eqn:fixed_pt_eqn} has at least $3$ solutions in $x$. From the symmetry in \eqref{eqn:fixed_pt_eqn}, since $x=0$ is always a root, it suffices to show the existence of at least one more root in $x>0$. First note the ranges of variables, $x \in (-\infty, \infty), \alpha \in (0, \infty), \beta \in (0, 1)$.
	
	Letting $r(x)$ be the right side of \eqref{eqn:fixed_pt_eqn}, since $0 \le Q(\cdot) \le 1$, we have
	\begin{align*}
		1-\beta \le r(x) := \frac{1 - \beta Q(-\alpha + \sigma_1x)}{1 - \beta Q(-\alpha - \sigma_1x)} \le \frac{1}{1-\beta},
	\end{align*}
	indicating that $r(x) \in [1-\beta, \tfrac{1}{1-\beta}]$. However, note that $e^x$ monotonically increases in $(1, \infty)$ for $x>0$. Since $e^x, r(x)$ coincide at $x=0$, it follows that they cross at least once on $(0, \infty)$ and also on $(-\infty, 0)$, if $r'(x) > \tfrac{d}{dx}e^x$ at $x=0$ by the intermediate value theorem. Thus, the sufficient condition follows:
	\begin{align*}
		r'(0) = \frac{2 \sigma_1\beta \phi(\alpha; 0, 1)}{1-\beta Q(-\alpha)} > 1 = \frac{d}{dx}e^x \Bigg|_{x=0}.
	\end{align*}
\end{IEEEproof}

\begin{figure}[t]
	\centering
	\includegraphics[width=3.5in]{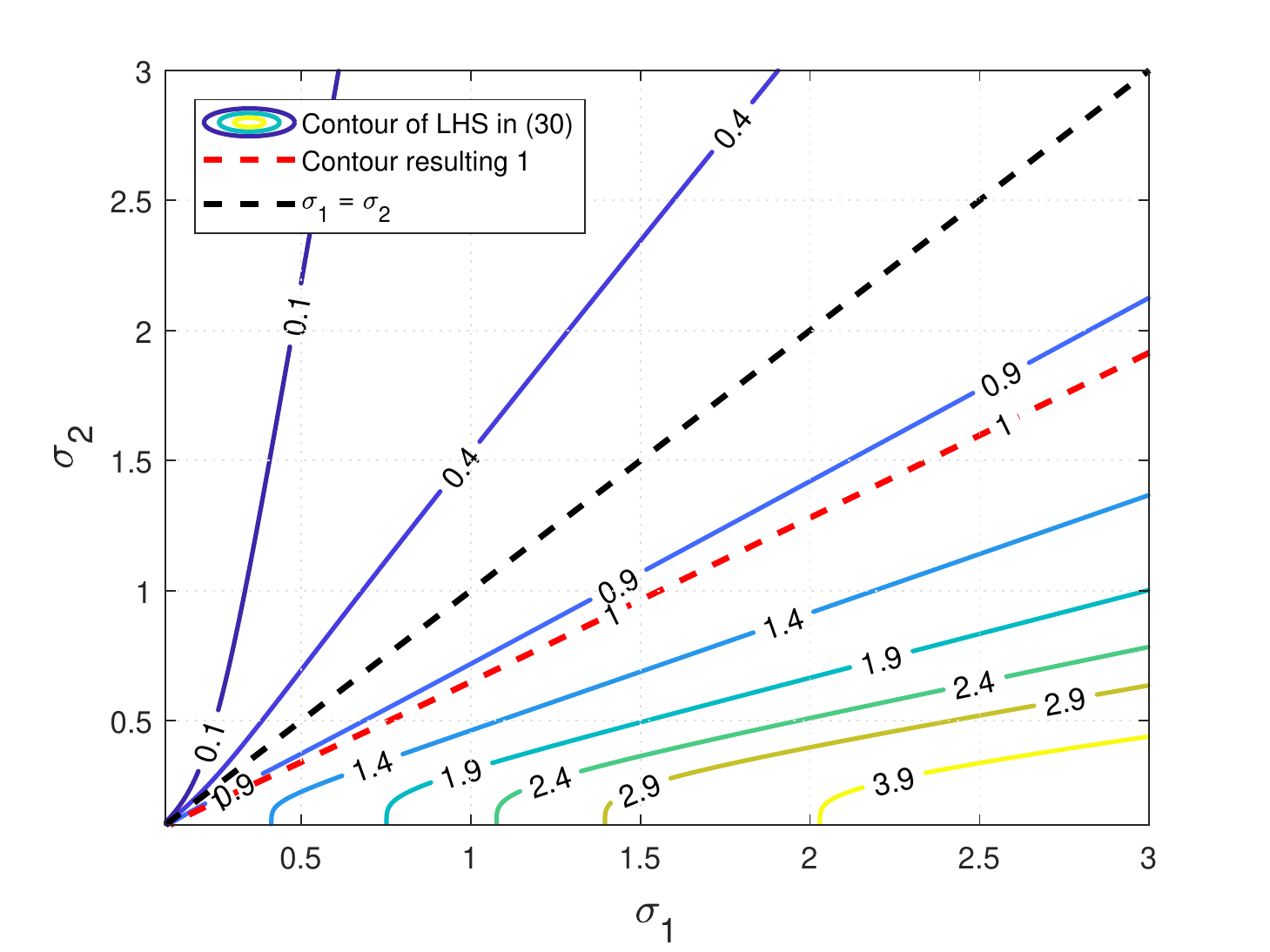}
	\caption{Contour plot of \eqref{eqn:suff_cond} with values for various $\sigma_1, \sigma_2$. The red dotted contour shows the contour that results in $1$ so that the area below it satisfies \eqref{eqn:suff_cond} and therefore has multiple solutions to \eqref{eqn:fixed_pt_eqn}. }
	\label{fig:suff_cond_noise}
\end{figure}

Cor.~\ref{cor:suff_cond} provides a sufficient condition on the noise level of agents under which there exists multiple crossings of the curves $q_1^*(p_0)$ and $p_0$. The range of standard deviations of the additive Gaussian noise of the preceding and last agents that satisfy the sufficient condition of Cor.~\ref{cor:suff_cond} is shown in Fig.~\ref{fig:suff_cond_noise}. Note from the figure that the area below the red dotted contour in Fig.~\ref{fig:suff_cond_noise} has multiple solutions to $q_1^* = p_0$, i.e., when the last agent has comparatively smaller than the preceding agent.

This is important as the crossings indicate a change in the perceived bias of the predecessor and also indicates the regions in which the last agent overweights the unlikely hypothesis as in Fig.~\ref{fig:sigma1>sigma2}.

\section{Team Construction Criterion} \label{sec:team_construction}

Having studied the mathematical conditions for optimal reweighting of initial beliefs, we now investigate team selection for social learning. Naturally, a social planner who is aware of the context $p_0$ can pick the optimal agent pairs to minimize Bayes risk. However, it is not clear if agents are capable of organizing themselves into ideal teams in the absence of contextual knowledge. Thus, we now identify the criterion for the last agent to identify the optimal predecessors among a set of given predecessors.

\begin{theorem} \label{thm:selection_cond}
Consider two predecessors with $q_1 < q_{1'}$. Let $\lambda_1, \lambda_{1'}$ be the decision thresholds of the respective predecessors. Then, the predecessor with belief $q_1$ is the optimal choice if and only if
\begin{equation} \label{eqn:selection_cond}
\frac{\Prob_1\qth{Y_1 \in [\lambda_1,\lambda_{1'}], Y_2 \in [\lambda_2^1,\lambda_2^0]}}{\Prob_0\qth{Y_1 \in [\lambda_1,\lambda_{1'}], Y_2 \in [\lambda_2^1,\lambda_2^0]}} \geq \frac{c_{10} p_0}{c_{01}(1-p_0)}.
\end{equation}
\end{theorem}
\begin{IEEEproof}
Given in App.~\ref{app:selection_cond}.
\end{IEEEproof}
In other words, by rewriting \eqref{eqn:selection_cond} in a likelihood ratio form, we observe that the criterion for picking the predecessor with a smaller belief is given by the likelihood ratio test
\[
\calL\qth{\wh{H}_1 = \wh{H}_2 = 1, \wh{H}_{1'} = \wh{H}_{2'} = 0} \geq \frac{c_{10} p_0}{c_{01}(1-p_0)},
\]
where $\wh{H}_{2'}$ is the decision made by the last agent following the decision of the predecessor with belief $q_{1'}$.

Thus selecting an ideal predecessor requires a social planner who is aware of the context $p_0$. Without this, the last agent selects an predecessor according to his personal belief $q_2$. That is, the last agent verifies condition \eqref{eqn:selection_cond} by replacing $p_0$ by $q_2$. Such a choice of predecessor might not always conform to the optimal choice when the belief of the last agent deviates significantly from the prior. To illustrate, we consider the problem of choosing between two predecessors with beliefs $q_1(p_0) = q_1^*(p_0)$ and $q_{1'}(p_0) = p_0$. Let $q(p_0,q_2)$ be the belief of the optimal predecessor choice for a given pair $(p_0,q_2)$. We identify the region of correct selection by shading, $\calS = \{(p_0,q_2): q(p_0,q_2) = q(q_2,q_2)\}$.

\begin{figure*}[t]
	\centering
	\subcaptionbox{
		Predecessor selection under equal noise. The optimal later-acting agent fails to recognize the optimal predecessor and makes mistakes often. The trend is similar in cases with smaller noise predecessor. \label{fig:equal_SNR_region}}[.32\textwidth][c]{
		\includegraphics[width=.32\textwidth]{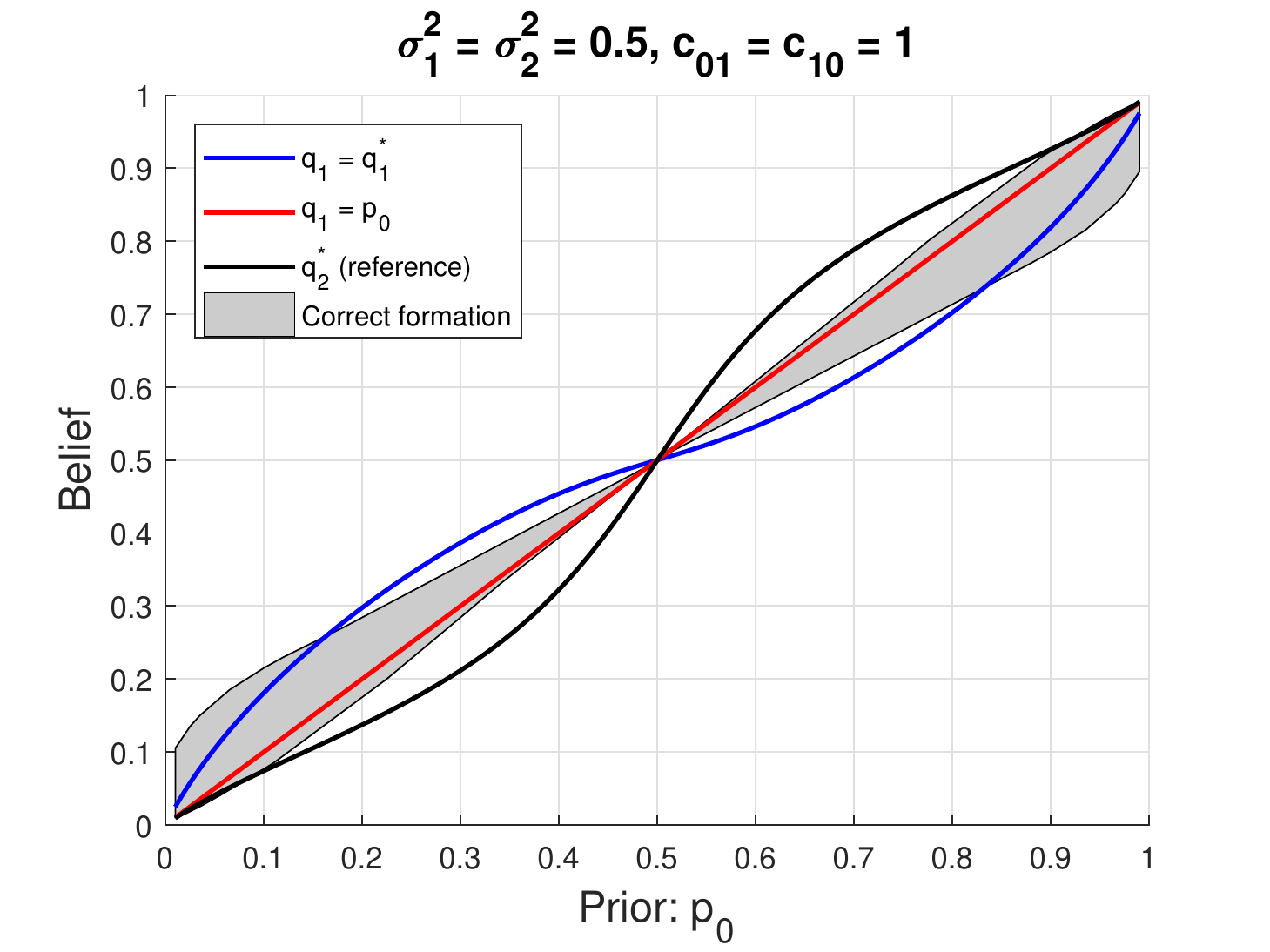}
	} ~
	\subcaptionbox{
		Predecessor selection under noise diversity. The optimal later-acting agent selects the optimal predecessor in this case. \label{fig:diff_SNR_region}}[.32\textwidth][c]{
		\includegraphics[width=.32\textwidth]{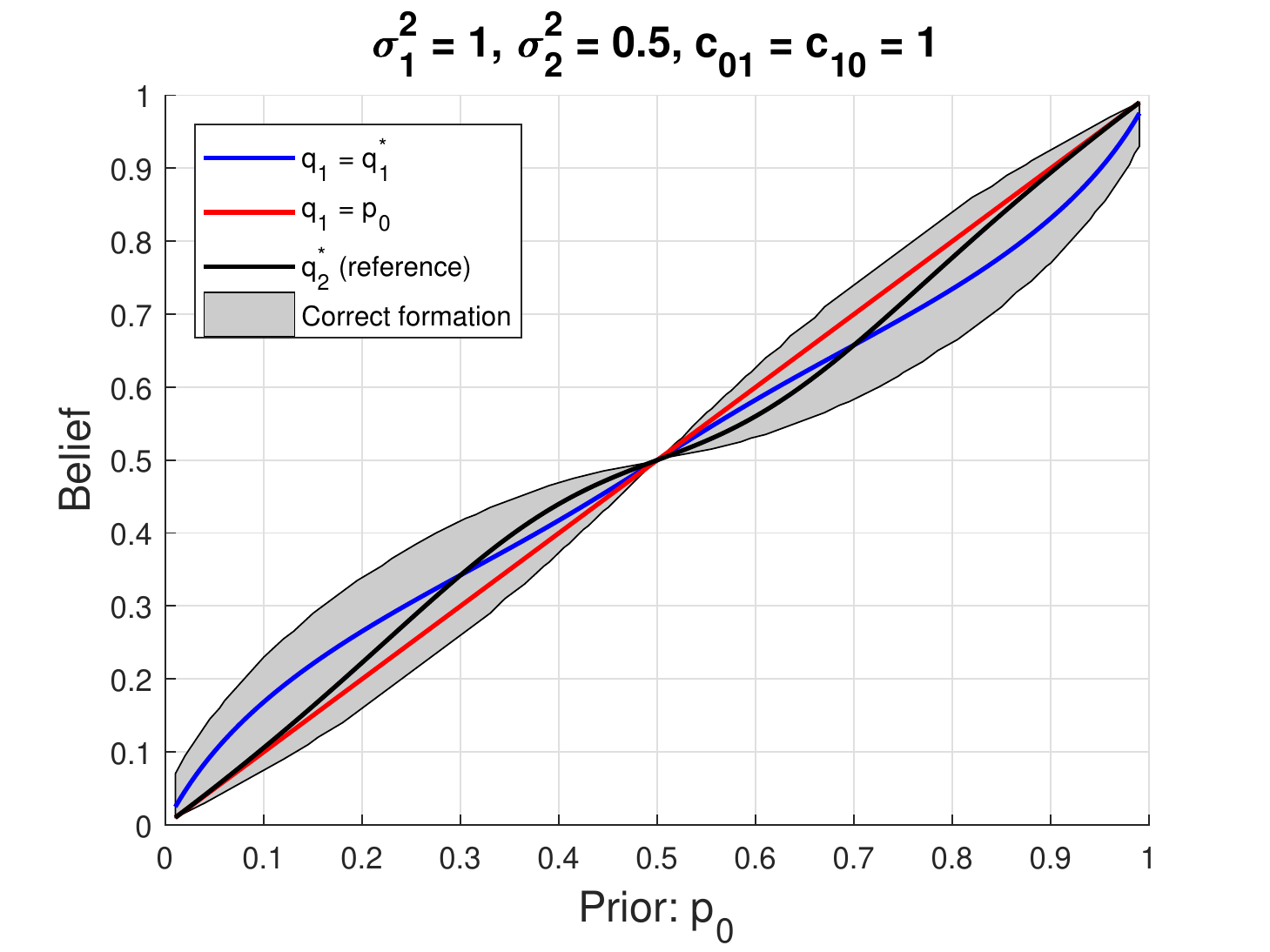}
	} ~
	\subcaptionbox{
		Context-unaware predecessor selection. The last agent chooses predecessor using \eqref{eqn:advisor_criterion_no_context} without $p_0$ and the Bayes risk increases as a result. \label{fig:suboptimal_choice}}[.32\textwidth][c]{
		\includegraphics[width=.32\textwidth]{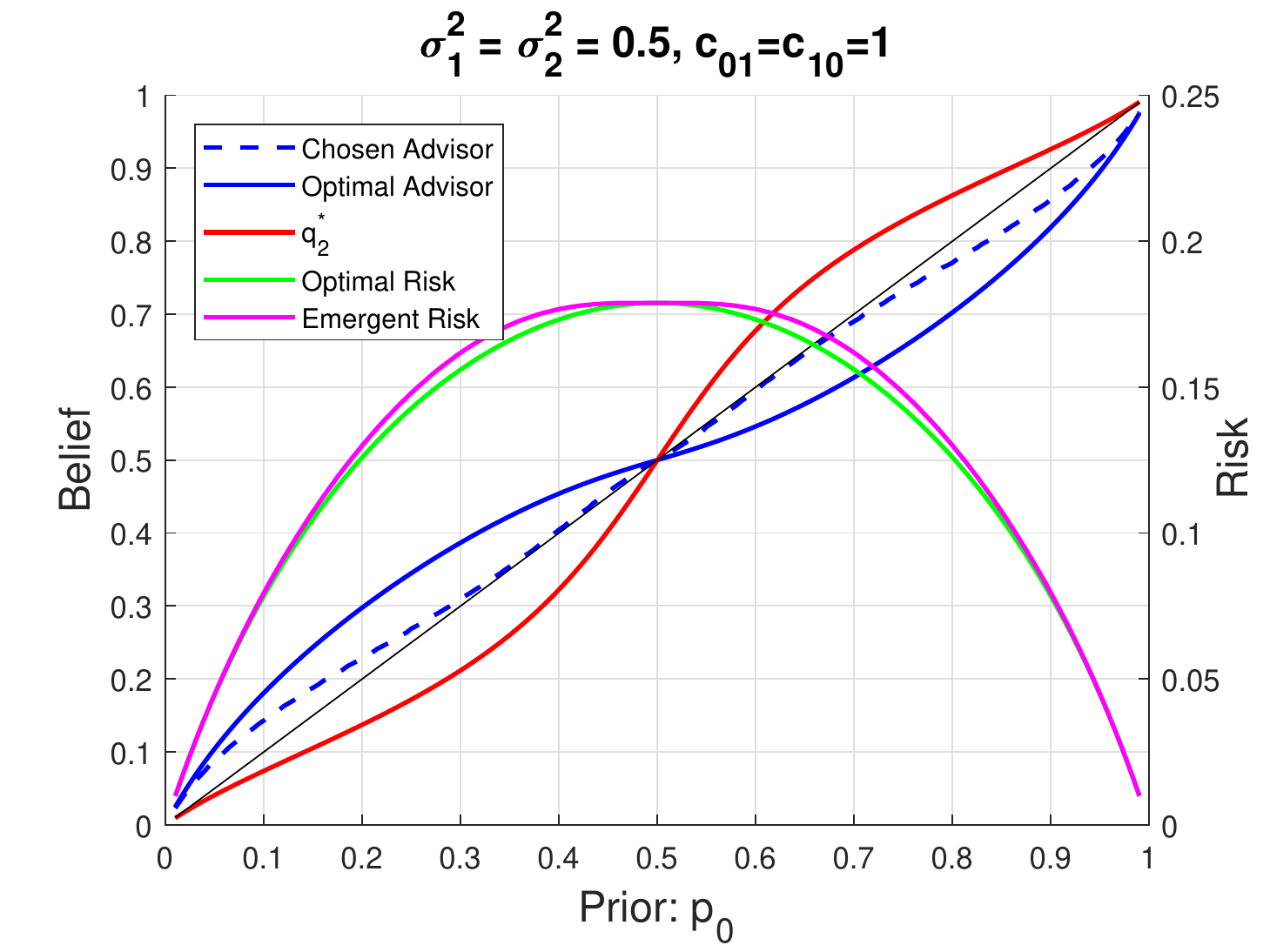}
	}
	\caption{Context-unaware team selection.}
\end{figure*}

First, when noise levels are equal, the region in which the last agent picks the correct preceding agent is shown in Fig.~\ref{fig:equal_SNR_region}. We note that the correct region is relatively small and does not include $q_2^*$. In particular, the last agent with optimal belief chooses the wrong predecessor always, whereas a suboptimal last agent with beliefs in the shaded region picks the correct one. 

On the other hand, when the last agent has smaller noise than the predecessor, the corresponding region is as shown in Fig.~\ref{fig:diff_SNR_region}. Here we note that the last agent with optimal belief picks the correct preceding agent always.

Thus, we note that knowledge of the mathematically optimal beliefs does not guarantee selection of the right preceding agent. Further, we also observe that the diversity of noise levels may increase the feasibility of selecting the right preceding agent when the last agent has optimal belief.

We also explore the optimal choice of predecessor for the given optimal last agent in the absence of knowledge of the prior probability. From \eqref{eqn:optimal_belief}, the belief of the optimal preceding agent, $\tilde{q}_1$ chosen by an last agent, in the absence of context (prior probability $p_0$) satisfies
\begin{equation} \label{eqn:advisor_criterion_no_context}
\frac{\tilde{q}_1}{1-\tilde{q}_1} = \frac{p_0}{1-p_0} \frac{P_{e,2}^{\RN{1}_1} - P_{e,2}^{\RN{1}_0}}{P_{e,2}^{\RN{2}_0} - P_{e,2}^{\RN{2}_1}}.
\end{equation}

The last agent's behavior with belief $q_2^*$ is as shown in Fig. \ref{fig:suboptimal_choice}. We note that the preceding agent chosen by the last agent differs from the optimal choice. Further, it is also evident that this choice consequently results in an increased Bayes risk. Such behavior in team selection highlights the significance of context and thus a social planner for identifying the right team.

\section{Human-AI Collaboration Systems} \label{sec:human_ai}

In this section, we use mathematical results from previous sections to study the engineering design problem of constructing human-AI collaborative systems. To do so, we make the following assumptions from the behavioral sciences: Human agents perform Bayesian decision-making \cite{SwetsTB1961, Viscusi1985, BraseCT1998, GlanzerHM2009} and their perceptions follow the Prelec reweighting function \cite{Prelec1998}. In addition, agents experience varying observational noise which is additive and Gaussian (as it is a common model in human signal perception \cite{KnillR1996, YostPF2003}). As usual in sequential social learning setup, all agents make selfish decisions \cite{BikhchandaniHW1998, GaleK2003}.

\subsection{Approximation by Prelec Family}

To design human-AI collaborative systems, we first determine whether optimal belief functions from previous sections are close to human behavior as modeled by cumulative prospect theory \cite{TverskyK1992, Prelec1998}.\footnote{Bounded rationality models have been categorized into two main classes---costly bounded rationality and truly bounded rationality [48]. Costly rationality considers the emergence of boundedly irrational behavior as optimization under some costs of decision-making such as computation and communication. On the other hand, truly bounded rationality is not based on an optimization framework. Though not the focus of the present paper, one might wonder whether people are (approximately) naturally optimal for social learning. That is, do cumulative prospect-theoretic models emerge from a costly rationality framework for social learning, since the optimal belief curves result from limitations in computation (selfish decision-making) and communication (public signal quantization). }

We approximate the optimal belief curves $q_n^*$ by the Prelec function and study the resulting increase in the Bayes risk. We restrict to the Prelec family whose fixed point is identical to $p^* = \frac{c_{10}}{c_{01}+c_{10}}$, and then find best parameters $(\alpha_n, \beta_n)$ in the minimax absolute error sense, i.e.,
\begin{align*}
(\alpha_n, \beta_n) = \argmin_{\alpha, \beta: w\pth{p^*;\alpha,\beta} = p^*} \| q_n^*(\cdot) - w(\cdot;\alpha, \beta) \|_{\infty}.
\end{align*}
Let the Prelec function approximations be $(q_{1, \textrm{Pre}}, q_{2, \textrm{Pre}})$.

\begin{figure}[t]
	\centering
	\begin{subfigure}[b]{0.23\textwidth}
		\includegraphics[width=1.7in]{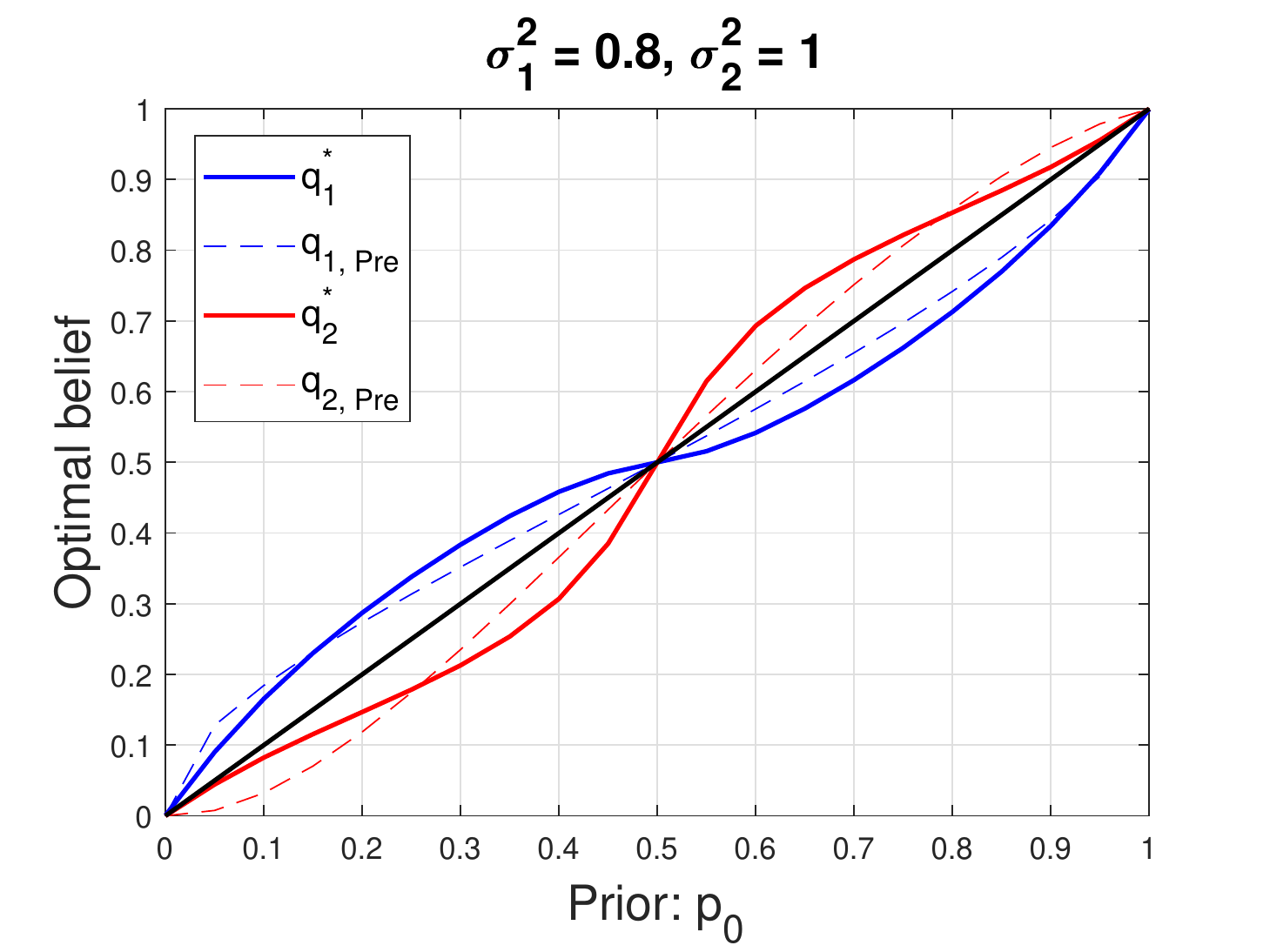}
		\subcaption{When the preceding agent has smaller noise.}
		\label{fig:sec7_prelec_approx_a}
	\end{subfigure}
	\begin{subfigure}[b]{0.23\textwidth}
		\includegraphics[width=1.7in]{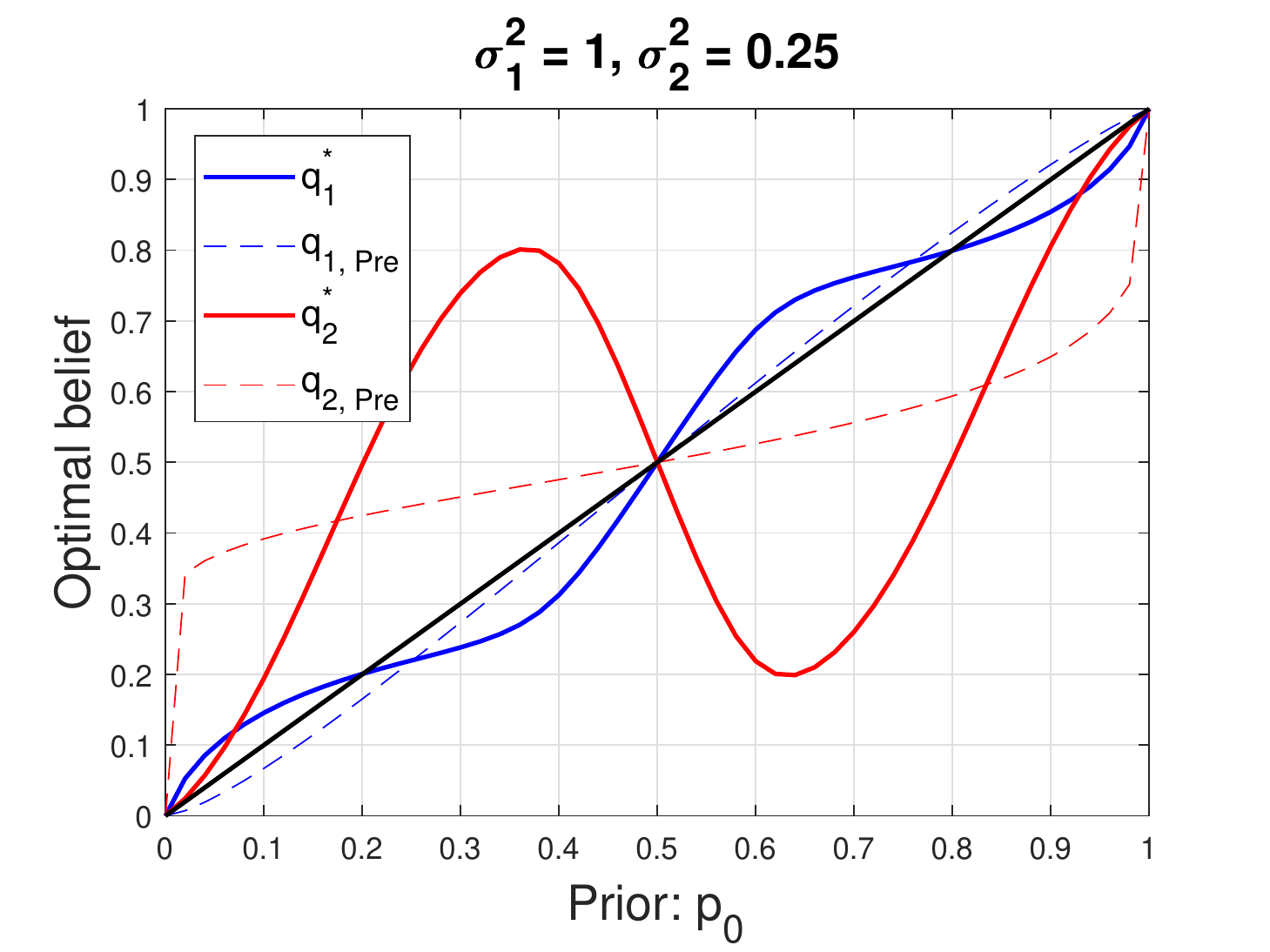}
		\subcaption{When the later-acting agent has smaller noise}
		\label{fig:sec7_prelec_approx_b}
	\end{subfigure}
	\caption{Optimal beliefs as compared to Prelec-weighted beliefs.}
	\label{fig:sec7_prelec_approx}
\end{figure}

The Prelec approximations for the two-agent case are shown in dotted curves in Fig.~\ref{fig:sec7_prelec_approx}. When the preceding agent has smaller noise as in Fig.~\ref{fig:sec7_prelec_approx_a}, the Prelec function approximates the optimal beliefs well and the Bayes risk does not increase by much. To evaluate the loss from the approximation, consider the set of correct beliefs $q_1=q_2=p_0$, that result in a Bayes risk of $R_{2,\textrm{corr}}$. The maximal loss in terms of Bayes risk from using the correct beliefs is $\max_{p_0} (R_{2,\textrm{corr}} - R_{2, \textrm{min}}) \approx 0.0039$. On the other hand, the maximal loss from the best Prelec approximation is $\approx 0.0009$. This indicates that the natural cognitive biases of humans (i.e., Prelec reweighting) are effective for social learning when the preceding agent has smaller noise.

On the other hand, when the last agent has smaller noise as in Fig.~\ref{fig:sec7_prelec_approx_b}, the Prelec approximation does not accurately mimic the optimal behavior of agents. Recall that the Prelec function is always increasing and has only one crossing with unit slope line in $(0,1)$. Therefore, the Prelec function fails to account for all the variations in the optimal belief. Moreover, while the additional loss of Bayes risk by the Prelec fitting is $\approx 0.0187$, the loss from using the correct beliefs, $p_0=q_1=q_2$, is $\approx 0.0060$. This indicates that even though the Prelec weighting functions serve as good approximations with predecessors having less noisy observations, they do not model the optimal behavior in the case of predecessors having noisier observations. These results suggest that human agents following cumulative prospect theory models \cite{TverskyK1992} yield small Bayes risk when predecessors have smaller noise.

\begin{figure}[t]
	\centering
	\begin{subfigure}[t]{0.5\textwidth}
		\centering
		\includegraphics[width=3.5in]{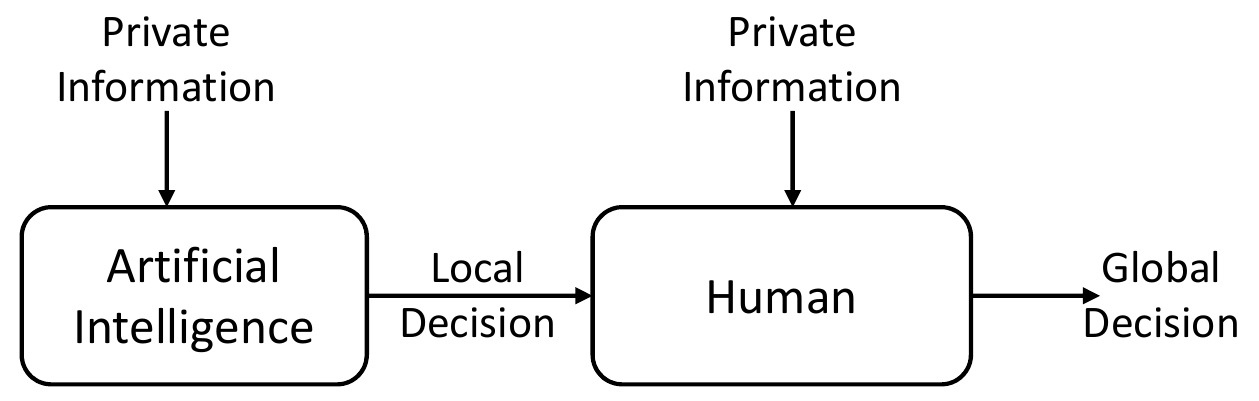}
		\caption{}
		\label{fig:AI_model_a}
	\end{subfigure}
	\\
	\begin{subfigure}[t]{0.5\textwidth}
		\centering
		\includegraphics[width=3.5in]{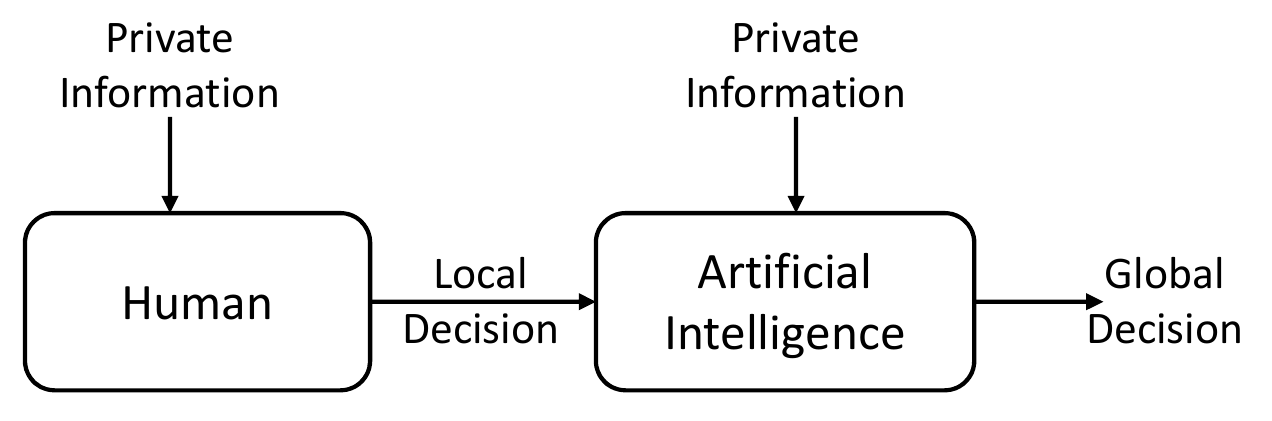}
		\caption{}
		\label{fig:AI_model_b}
	\end{subfigure}
	\caption{Models of AI-human collaboration, where a machine provides input for human judgement or vice versa.}
\end{figure}

\subsection{Human-AI Teams}
The previous subsection informs the design of AI-human collaboration structures \cite{McAfeeB2017}. In many human-AI joint teams, a human agent makes the final decision based on the advice of an AI component as depicted in Fig.~\ref{fig:AI_model_a}, but the opposite structure of Fig.~\ref{fig:AI_model_b} is also possible. It is thus important to identify the best team configuration \cite{Jarrahi2018}. Indeed, D.~Kahneman recently stated that ``You can combine humans and machines, provided the machine has the last word'' \cite{MITIDE2018}.

Our results indicate that an AI assistant with smaller noise could be an effective predecessor to the human decision-maker. In particular, an open-minded AI predecessor and a closed-minded human final decision-maker with appropriate Prelec reweighted beliefs work well together, as in Fig.~\ref{fig:sigma1<sigma2}. However, an AI component with greater noise might not be a good predecessor to the human last agent who does not have beliefs that mimic the optimal behavior in Fig.~\ref{fig:sigma1>sigma2} and so perhaps counterintuitively, the architecture of Fig.~\ref{fig:AI_model_b} should be adopted, with the AI agent having larger noise making the global decision.

Additionally, these results along with those of Sec.~\ref{sec:team_construction} provide some insight into human-AI teams when the human agent picks an AI predecessor, given a choice among different agents. In particular, consider the AI-human team where the human, who has a Prelec-weighted belief, chooses one of two possible AI predecessors---one that has the optimal belief $q_1^*$ and the other that is aware of the true prior $p_0$. In case the human agent has larger noise, and a closed-minded Prelec belief as in Fig.~\ref{fig:equal_SNR_region}, she unfortunately picks the AI predecessor with $q_1 = p_0$ and the team becomes suboptimal. However, if the human agent has smaller noise, and an open-minded Prelec belief, she picks the optimal AI component $q_1 = q_1^*$ and therefore can make the optimal decision as in Fig.~\ref{fig:diff_SNR_region}. Thus it is evident that optimal team organization is feasible when the human has smaller noise and the appropriate open-minded belief.

\section{Conclusion} \label{sec:conclusion}

We discussed the sequential social learning problem with individual biased beliefs. Unlike previous works on herding, we focused on the Bayes risk of the last-acting agent. We first derived the optimal belief update rule for general likelihoods and evaluated for Gaussian likelihoods. Counterintuitively, optimal beliefs that yield minimum Bayes risk are in general different from the true prior. Under equal noise levels, we observed that optimal preceding agents have open-minded beliefs, that is, overweight small priors and underweight large priors, while the optimal last agent has closed-minded belief. However, the trend may change depending on varying noise levels such that especially when the last agent has much smaller noise, optimal belief of the last agent is inverted as she becomes open-minded.

We also showed that the Prelec reweighting function from cumulative prospect theory approximates the behavior of the optimal beliefs under specific levels of noise, however, when the last agent has much smaller noise, it fails to capture all the behavioral traits of the optimal beliefs.

Finally, we considered the ability of agents to organize themselves into optimal teams and showed that in the absence of a social planner, the last agent can get paired with the wrong predecessor when the individual belief deviates significantly from the underlying prior value. The setup arises from the consideration of AI and it tells us without knowing the true prior, our human-machine team construction could be misorganized.

\appendices
\section{Proof of Theorem \ref{prop:belief_props}} \label{app:belief_props}
Let us prove Thm.~\ref{prop:belief_props} starting with the premise that $q_1^* \geq p_0$. First, from \eqref{eqn:optimal_belief}, we have
\begin{align}
q_1^* \geq p_0 &\iff \frac{P_{e,2}^{\RN{2}_1} - P_{e,2}^{\RN{2}_0}}{P_{e,2}^{\RN{1}_1} - P_{e,2}^{\RN{1}_0}} \geq -1. \label{eqn:slope_condn}
\end{align}

\begin{figure}[t]
	\centering
	\includegraphics[width=3.5in]{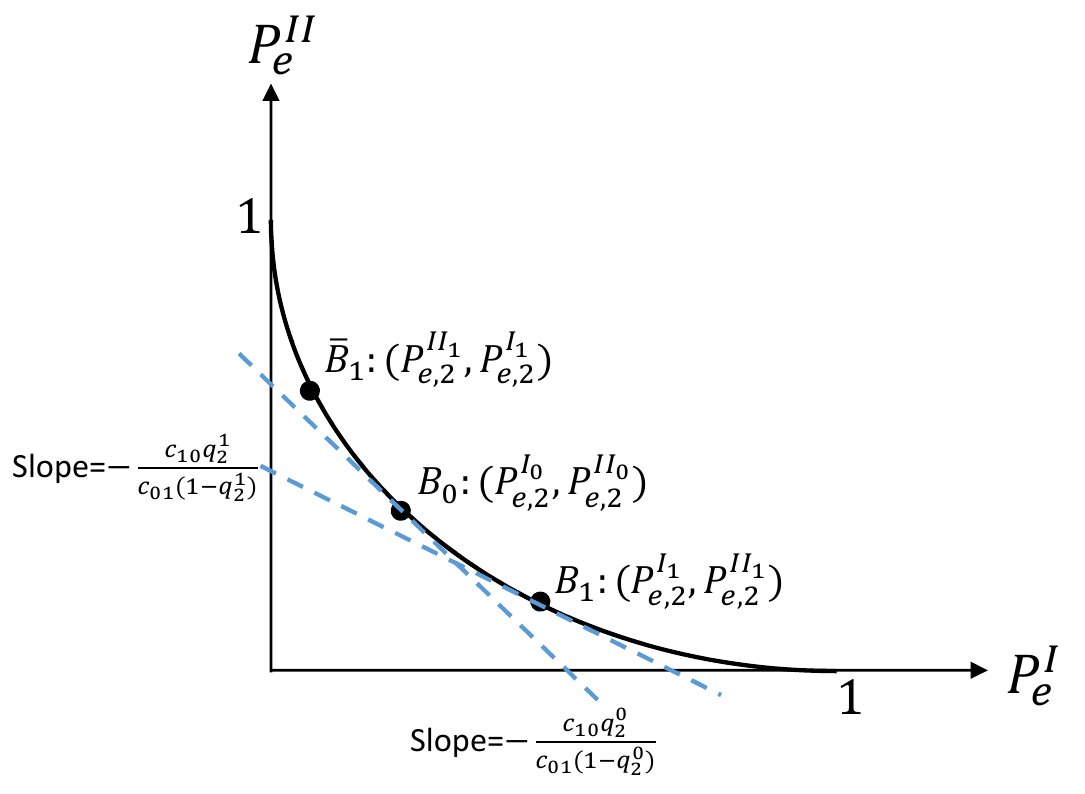}
	\caption{The point $B_0$ always exists between points $B_1$ and $\bar{B}_1$.}
	\label{fig:error_curve}
\end{figure}

To study the ratio in \eqref{eqn:slope_condn}, consider the Type I vs. Type II error curve for binary hypothesis testing under additive Gaussian noise.\footnote{It is also called Receiver Operating Characteristic (ROC) curve \cite{VanTrees1968, Poor1988} when the curve is vertically inverted.} This is shown in Fig.~\ref{fig:error_curve}, and as seen here is a convex function \cite{VanTrees1968}. Note that on the curve, the Type I and Type II error probabilities, $(P_e^{\RN{1}}, P_e^{\RN{2}})$, are the points on the curve that have tangents with slope matching $-\pth{\tfrac{c_{10}q}{c_{01}(1-q)}}$, where $q$ is the corresponding probability, and $\sigma^2$ is the variance of the additive Gaussian noise. 

First, from Thm.~\ref{thm:order_preserve1}, we know that $q_2^0 \geq q_2^1$ which in turn implies that $\lambda_2^0 \geq \lambda_2^1$. This in turn indicates that 
\[
P_{e,2}^{\RN{1}_0} = Q\pth{\tfrac{\lambda_2^0}{\sigma_2}} \leq Q\pth{\tfrac{\lambda_2^1}{\sigma_2}} = P_{e,2}^{\RN{1}_1}.
\]
Similarly, $P_{e,2}^{\RN{2}_0} \geq P_{e,2}^{\RN{2}_1}$, and thus, as shown in the figure, the point $B_0 = \pth{P_{e,2}^{\RN{1}_0},P_{e,2}^{\RN{2}_0}}$ lies to the left of $B_1 = \pth{P_{e,2}^{\RN{1}_1},P_{e,2}^{\RN{2}_1}}$.

Further, since $B_1$ lies on the curve, so does the point $\bar{B}_1 = \pth{P_{e,2}^{\RN{2}_1},P_{e,2}^{\RN{1}_1}}$ as it caters to the error probabilities corresponding to the probability of the null hypothesis $\prob{H=0} = 1-q_2^1$. Thus, the line $\overline{B_1\bar{B}_1}$ has a slope of $-1$. 

Note that the condition \eqref{eqn:slope_condn} translates to the slope of the line $\overline{B_0B_1}$ is greater than $-1$. Observe that if $\bar{B}_1$ lies to the right of $B_1$ then it implies that the slope of $\overline{B_0B_1}$ is less than $-1$, violating \eqref{eqn:slope_condn}.  Similarly, if $B_0$ lies to the left of $\bar{B}_1$, then again the \eqref{eqn:slope_condn} is violated. 

On the other hand, if $B_0$ lies between $\bar{B}_1$ and $B_1$, then we know that the slope of $\overline{B_0B_1}$ is greater than that of $\overline{B_1\bar{B}_1}$, therein satisfying \eqref{eqn:slope_condn}. Thus, \eqref{eqn:slope_condn} is true if and only if the point $B_0$ lies between the two points $B_1$ and $\bar{B}_1$.

From the convexity of the curve and comparing coordinates of $B_0$ and $\bar{B}_1$, we have
\begin{align}
&q_1^* \geq p_0 \nonumber \\
&\iff P_{e,2}^{\RN{1}_0} \geq P_{e,2}^{\RN{2}_1} \text{ and } P_{e,2}^{\RN{2}_0} \leq P_{e,2}^{\RN{1}_1} \notag \\
&\stackrel{(a)}{\iff} Q\pth{\frac{\lambda_2^0}{\sigma_2}} \geq 1 - Q\pth{\frac{\lambda_2^1-1}{\sigma_2}} \nonumber \\
&~~~ \text{ and } Q\pth{\frac{\lambda_2^1}{\sigma_2}} \geq 1 - Q\pth{\frac{\lambda_2^0-1}{\sigma_2}} \nonumber \\
&\stackrel{(b)}{\iff} \lambda_2^0 + \lambda_2^1 \leq 1 \nonumber \\
&\stackrel{(c)}{\iff} 2\lambda_{1,[2]} + \sigma_2^2 \log\pth{\frac{P_{e,1,[2]}^{\RN{1}}\pth{1- P_{e,1,[2]}^{\RN{1}}}}{P_{e,1,[2]}^{\RN{2}}\pth{1- P_{e,1,[2]}^{\RN{2}}}}} \leq 1, \label{eqn:condn_thresh_plugin}
\end{align}
where (a) follows from the false alarm and missed detection probabilities in terms of the $Q$-function of the standard Gaussian random variable; (b) follows from the fact that the $Q$-function is monotonically decreasing and that $1 - Q(x) = Q(-x)$; and (c) follows from \eqref{eqn:q20_posterior}, \eqref{eqn:q21_posterior}, and $\lambda_{1, [2]} = \lambda_2(q_2)$.

From \eqref{eqn:Gauss_BHT_threshold}, we have
\[
\lambda_{1,[2]} = \frac{1}{2} + \sigma_2^2 \log \pth{\frac{c_{10}q_2^*}{c_{01}(1-q_2^*)}}.
\]
Substituting in \eqref{eqn:condn_thresh_plugin}, we have
\begin{align*}
q_1^* \geq p_0 &\iff 2 \log \pth{\frac{c_{10}q_2^*}{c_{01}(1-q_2^*)}} \\
&~~~ ~~~ \leq \log\pth{\frac{P_{e,1,[2]}^{\RN{2}}\pth{1- P_{e,1,[2]}^{\RN{2}}}}{P_{e,1,[2]}^{\RN{1}}\pth{1- P_{e,1,[2]}^{\RN{1}}}}}.
\end{align*}

Letting $x := \log \pth{\tfrac{c_{10}q_2^*}{c_{01}(1-q_2^*)}} = \tfrac{1}{\sigma_2^2} \pth{\lambda_2 - \tfrac{1}{2}}$ and using $Q(\cdot)$ representation of error probabilities, we have
\begin{align}
q_1^* \geq p_0 &\iff 2x \leq \log \pth{\frac{Q\pth{\sigma_2 x - \tfrac{1}{2\sigma_2}} Q\pth{-\sigma_2 x + \tfrac{1}{2\sigma_2}} }{Q\pth{\sigma_2 x + \tfrac{1}{2\sigma_2}} Q\pth{-\sigma_2 x - \tfrac{1}{2\sigma_2}} }}. \label{eqn:Q_func_condn}
\end{align}
From Cor.~\ref{lem:inc_fn_gauss}, we know that the function
\[
\tilde{g}(x) = x + \log \pth{\frac{Q\pth{\sigma x + \tfrac{1}{2\sigma}}}{Q\pth{\sigma x - \tfrac{1}{2\sigma}}}}
\]
is an increasing function of $x$. Thus, reformulating \eqref{eqn:Q_func_condn} using $\tilde{g}(\cdot)$, 
\begin{align*}
q_1^* \geq p_0 &\iff \tilde{g}(x) \leq \tilde{g}(-x) \\
&\iff x \leq 0 \iff q_2^* \leq \frac{c_{01}}{c_{01}+c_{10}}.
\end{align*}
The condition for equality follows from observing the condition for equality at all the inequalities, proving the first part of the result.

The second part follows directly from the first, taking into account the trivial cases of $p_0 \in \{0,1\}$.

\section{Proof of Theorem \ref{thm:fixed_pt}} \label{app:fixed_pt}
We will consider the case of $c_{01} = c_{10} = 1$ for convenience. The proof extends directly by a simple scaling argument.

The optimal belief of worker two satisfies $\frac{\partial R_2}{\partial q_2} = 0$. Thus, differentiating \eqref{eqn:Blake_Bayes_risk} with respect to $q_2$ and rearranging,
\begin{align*}
& p_0\qth{(1-P_{e,1}^{\RN{1}})f_{Y_2 \vert H}(\lambda_2^0\vert 0) \frac{\partial \lambda_2^0}{\partial q_2} + P_{e,1}^{\RN{1}} f_{Y_2 \vert H}(\lambda_2^1\vert 0) \frac{\partial \lambda_2^1}{\partial q_2}} = \\
& (1-p_0)\qth{P_{e,1}^{\RN{2}}f_{Y_2 \vert H}(\lambda_2^0\vert 1) \frac{\partial \lambda_2^0}{\partial q_2} + (1-P_{e,1}^{\RN{2}}) f_{Y_2 \vert H}(\lambda_2^1\vert 1) \frac{\partial \lambda_2^1}{\partial q_2}}.
\end{align*}

Let $x = \log\pth{\frac{p_0}{1-p_0}}$. For $q_2^* = 1/2$ and $q_1^* = p_0$, we have 
\begin{align*}
\lambda_1 = \frac{1}{2} + \sigma_1^2 x \textrm{ and } \lambda_{1,[2]} = \frac{1}{2}.
\end{align*} 
It implies $P_{e,1,[2]}^{\RN{1}} = P_{e,1,[2]}^{\RN{2}} = Q(1/2\sigma_2)$. Then,
\begin{align*}
\calL(\lambda_2^0) &= \frac{f_{Y_2\vert H}(\lambda_2^0 \vert 1)}{f_{Y_2\vert H}(\lambda_2^0 \vert 0)} \\
&= \frac{q_2}{1-q_2}\frac{(1-P_{e,1,[2]}^{\RN{1}})}{P_{e,1,[2]}^{\RN{2}}} = \frac{Q(-1/2\sigma_2)}{Q(1/2\sigma_2)} =: \frac{1}{c}, \\
\calL(\lambda_2^1) &= \frac{f_{Y_2\vert H}(\lambda_2^1 \vert 1)}{f_{Y_2\vert H}(\lambda_2^1 \vert 0)} \\
&= \frac{q_2}{1-q_2}\frac{P_{e,1,[2]}^{\RN{1}}}{(1-P_{e,1,[2]}^{\RN{2}})} = \frac{Q(1/2\sigma_2)}{Q(-1/2\sigma_2)} = c.
\end{align*}
Equivalently, this implies that
\[
\lambda_2^0 = \frac{1}{2} + \sigma^2 \log\pth{\frac{1}{c}}, \quad \lambda_2^1 = \frac{1}{2} - \sigma^2 \log\pth{\frac{1}{c}}.
\]
Thus, $\lambda_2^0 + \lambda_2^1 = 1$, and so,
\begin{align*}
f_{Y_2\vert H}(\lambda_2^1 \vert 1) &= \frac{1}{\sqrt{2 \pi}\sigma_2} \exp\pth{-\tfrac{(\lambda_2^1 - 1)^2}{2\sigma_2^2}} \\
&= \frac{1}{\sqrt{2 \pi}\sigma_2}\exp\pth{-\tfrac{(\lambda_2^0)^2}{2\sigma_2^2}} \\
&= f_{Y_2\vert H}(\lambda_2^0 \vert 0).
\end{align*}
Similarly, we also have
\[
f_{Y_2\vert H}(\lambda_2^1 \vert 0) = f_{Y_2\vert H}(\lambda_2^0 \vert 1).
\]

Further, from \eqref{eqn:q20_posterior} and \eqref{eqn:q21_posterior}, we have
\begin{align*}
\frac{d \lambda_2^0}{dq_2} &= \frac{d \lambda_2^0}{d \lambda_{1,[2]}} \frac{d \lambda_{1,[2]}}{dq_2} \\
&= \qth{1 + \frac{\sigma_2^2 \phi\pth{\tfrac{\lambda_{1,[2]}}{\sigma_2}}}{1-P_{e,1,[2]}^{\RN{1}}} - \frac{\sigma_2^2 \phi\pth{\tfrac{\lambda_{1,[2]}-1}{\sigma_2}}}{P_{e,1,[2]}^{\RN{2}}} }\frac{d\lambda_{1,[2]}}{dq_2}, \\
\frac{d \lambda_2^1}{dq_2} &= \frac{d \lambda_2^1}{d \lambda_{1,[2]}} \frac{d \lambda_{1,[2]}}{dq_2} \\
&= \qth{1 - \frac{\sigma_2^2 \phi\pth{\tfrac{\lambda_{1,[2]}}{\sigma_2}}}{P_{e,1,[2]}^{\RN{1}}} + \frac{\sigma_2^2 \phi\pth{\tfrac{\lambda_{1,[2]}-1}{\sigma_2}}}{1 - P_{e,1,[2]}^{\RN{2}}} }\frac{d\lambda_{1,[2]}}{dq_2}.
\end{align*}
When, $\lambda_{1,[2]} = \tfrac{1}{2}$, $P_{e,1,[2]}^{\RN{1}} = P_{e,1,[2]}^{\RN{2}} = Q\pth{\tfrac{1}{2\sigma_2}}$, and $\phi\pth{\tfrac{\lambda_{1,[2]}}{\sigma_2}} = \phi\pth{\tfrac{\lambda_{1,[2]}-1}{\sigma_2}}$. Thus, $\tfrac{d \lambda_2^0}{dq_2} = \tfrac{d \lambda_2^1}{dq_2}$.

Using these, the values of prior for which $q_1^* = p_0, q_2^* = 1/2$ are given by
\begin{align}
\frac{p_0}{1-p_0} = \frac{Q\pth{\tfrac{-1}{2\sigma_2}}Q\pth{\tfrac{-1}{2\sigma_1} - \sigma_1 x} + Q\pth{\tfrac{1}{2\sigma_2}}Q\pth{\tfrac{1}{2\sigma_1} + \sigma_1x}}{Q\pth{\tfrac{-1}{2\sigma_2}}Q\pth{\tfrac{-1}{2\sigma_2} +\sigma_1 x} + Q\pth{\tfrac{1}{2\sigma_2}}Q\pth{\tfrac{1}{2\sigma_1} - \sigma_1x}}. \label{eqn:full_form}
\end{align}
Using the definitions of $x, \alpha, \beta$ in \eqref{eqn:full_form}, and the fact that $Q(-y) = 1-Q(y)$, the result follows.

\section{Proof of Theorem \ref{thm:selection_cond}} \label{app:selection_cond}
From \eqref{eqn:Bayes_risk_n}, we note that the Bayes risk for social learning with beliefs $(q_1,q_2)$ is
\begin{align*}
R_2(q_1,q_2) &= c_{10} p_0 \qth{P_{e,2}^{\RN{1}_0}(1-P_{e,1}^{\RN{1}}) + P_{e,2}^{\RN{1}_1}P_{e,1}^{\RN{1}}} \\
&~~~ + c_{01} (1-p_0) \qth{P_{e,2}^{\RN{2}_0}P_{e,1}^{\RN{2}} + P_{e,2}^{\RN{2}_1}(1-P_{e,1}^{\RN{2}})}.
\end{align*}
Then, the difference in Bayes risk between the two choices of advisors is given by
\begin{align}
\Delta R_2 &= R_2(q_1,q_2) - R_2(q_{1'},q_2) \notag \\
&= c_{10}p_0 (P_{e,1}^{\RN{1}} - P_{e,1'}^{\RN{1}})(P_{e,2}^{\RN{1}_1} - P_{e,2}^{\RN{1}_0}) \nonumber \\
&~~~ + c_{01}(1-p_0)(P_{e,1}^{\RN{2}} - P_{e,1'}^{\RN{2}})(P_{e,2}^{\RN{2}_0} - P_{e,2}^{\RN{2}_1}). \label{eqn:delta_R}
\end{align}

Since $q_1 < q_{1'}$, the decision thresholds satisfy $\lambda_1 < \lambda_{1'}$. Thus, from \eqref{eqn:delta_R} and independence of $Y_1, Y_2$ given $H$, we see that $\Delta R_2 \leq 0$ if and only if \eqref{eqn:selection_cond} holds.

\end{document}